\documentclass[aps, prb, reprint, twocolumn,floatfix, superscriptaddress]{revtex4-2}
\usepackage{graphicx} 
\usepackage{xcolor}

\usepackage{dblfloatfix}

\usepackage{dcolumn}
\usepackage{bm}
\usepackage[version=4]{mhchem}

\begin{document}


\title{Defective $\alpha$-In$_2$Se$_3$ from first principles}

\author{Jorge Vega Bazantes}
\affiliation{Department of Physics and Engineering Physics, Tulane University, New Orleans, Louisiana 70118, USA}

\author{Timo Lebeda}
\affiliation{Yusuf Hamied Department of Chemistry, University of Cambridge, Cambridge CB2 1EW, UK}

\author{Zhiqiang Mao}
\affiliation{Department of Physics, The Pennsylvania State University, University Park, Pennsylvania 16802, USA}

\author{Jianwei Sun}
\thanks{Contact author: jsun@tulane.edu}
\affiliation{Department of Physics and Engineering Physics, Tulane University, New Orleans, Louisiana 70118, USA}

\date{November 10, 2026}

\begin{abstract}
$\alpha$-In$_2$Se$_3$ is a two-dimensional van der Waals material with promising applications involving ferroelectricity. However, the presence of native defects can significantly influence its structural and electronic properties. In this work, we employ density functional theory with a state-of-the-art meta-generalized gradient approximation (meta-GGA) exchange-correlation functional to investigate the electronic structure of bulk $\alpha$-In$_2$Se$_3$ and the formation of neutral point defects, including vacancies and antisites, under Se-rich and In-rich growth conditions. Our calculations reproduce the experimental structural parameters of the pristine material and reveal that the Se vacancy in the middle layer exhibits an unexpectedly low formation energy under both chemical environments. Structural analysis shows that the neighboring tetrahedrally coordinated In atom undergoes a substantial local relaxation along the three-fold axis, which preserves its $C_{3v}$ site symmetry even as its coordination is reduced from four to three, providing a microscopic explanation for the remarkable stability of this defect.

\end{abstract}


\maketitle


\section{\label{sec:level1}INTRODUCTION}
The study of 2D materials over the last decades has paved the way for the discovery of a myriad of compounds with applications in computing and semiconductor devices \cite{liu2020two}, photonics \cite{xia2014two}, quantum information \cite{liu20192d}, and ferroelectrics \cite{wang2023towards} due to their unique electronic and structural properties. An exemplary compound of this class is In$_2$Se$_3$, which has become a promising candidate for several applications in optoelectronic \cite{shi2020ultrafast,xue2020optoelectronic}, photonic \cite{kou2020catalyst,jacobs2014extraordinary}, thermoelectric \cite{cui2015high,cui2017significantly} and semiconductor and memory devices \cite{huang2022two, yan2022giant}. The promising range of potential applications of In$_2$Se$_3$ arises from its distinctive and complex atomic structure. In fact, several phases of In$_2$Se$_3$ have been identified, with the most common regarded as $\alpha$, $\beta$ and $\gamma$ \cite{kupers2018controlled}. 
In addition, early experimental studies investigated the phase behavior and crystal structures of In$_2$Se$_3$ \cite{miyazawa1957,osamura1966crystal,popovic1979revised,likforman1980transitions}. More recent diffraction, microscopy, and first-principles work clarified the layered structure and polytypes of $\alpha$-In$_2$Se$_3$ \cite{kupers2018controlled}. Building on this structural picture, Ding et al.\ predicted coupled in-plane and out-of-plane ferroelectric polarization in a single $\alpha$ In$_2$Se$_3$ layer \cite{ding2017prediction}. Zhou et al.\ subsequently reported experimental evidence for out-of-plane polar order in $\alpha$-In$_2$Se$_3$ nanoflakes \cite{zhou2017out}, while Xiao et al.\ provided further evidence for electrically switchable polarization and dipole locking \cite{xiao2018intrinsic}.

Moreover, the specific synthesis route employed to fabricate a two-dimensional (2D) material can introduce structural defects, which may either degrade or improve the resulting material properties. For example, point defects introduced during growth or irradiation can modify the electronic and ferroelectric properties of $\alpha$-In$_2$Se$_3$. Hou et al.\ combined neutron-irradiation experiments with PBE \cite{perdew1996generalized} calculations of the effects of In vacancies on electronic structure and polarization \cite{hou2020neutron}. Regarding an $\alpha$-In$_2$Se$_3$ monolayer, Tang et al.\ calculated vacancy formation energies and examined defect-controlled polarization and doping, while Ma et al.\ investigated the coupling between vacancies, polarization, and conductivity \cite{tang2021controllable,ma2025vacancy}. More recently, Pang et al.\ combined scanning tunneling microscopy and spectroscopy of bulk crystals with PBE-based defect modeling to identify In vacancies and In antisites \cite{pang2025scanning}. These studies establish the importance of the study of defects for $\alpha$-In$_2$Se$_3$.

On the other hand, density functional theory (DFT) is a fundamental tool for studying layered materials that exhibit van der Waals (vdW) interactions. To account for vdW interactions, which are non-local, meta-GGA functionals such as SCAN \cite{scan,sun2016accurate} and its regularized form $r^2$SCAN \cite{furness2020accurate} have been combined with the rVV10 non-local correlation to give SCAN+rVV10 \cite{peng2016versatile} and $r^2$SCAN+rVV10 \cite{ning2022workhorse}, providing an improved balance between accuracy and computational efficiency. In particular, in semiconducting materials meta-GGAs can be expected to provide a more accurate description of the defect formation energy than GGAs \cite{bazantes2026rationalizing}.

In this work we study the $\alpha$ phase of In$_2$Se$_3$ with 3R stacking at the meta-GGA level using $r^2$SCAN+rVV10, analyzing the structural and electronic properties of the bulk and the thermodynamic stability of neutral point defects, including vacancies and antisites. Our central result is that the Se vacancy in the middle layer, $V_{Se2}$, is by about $1$~eV the most stable of the three inequivalent Se vacancies, and remains competitive with the $Se_{In}$ antisites even under Se-rich conditions. Through a bond-counting and bond-valence analysis, and by comparing relaxed with unrelaxed defect formation energies, we rationalize why $V_{Se2}$ is the most stable Se vacancy and trace its stability to a symmetry-allowed relaxation.

\section{Computational Methods}\label{sec:comp_methods}
Density functional theory (DFT) calculations were performed using the plane wave Vienna $ab$ $initio$ simulation package (VASP) \cite{kresse1996efficient,kresse1993ab} implementing the pseudopotential projector-augmented wave method (PAW) \cite{kresse1999ultrasoft}.
We use the meta-GGA $r^2$SCAN+rVV10 \cite{ning2022workhorse} for all calculations.
We first relax the primitive bulk structure of $\alpha$-In$_{2}$Se$_{3}$ with 3R stacking, which belongs to the space group R3$m$ (160), using a regular $k$-point mesh of $11\times11\times4$ on the primitive cell, which contains one formula unit and has $|a_{1}|=|a_{2}|=4.04$~\AA\ and $|a_{3}|=9.95$~\AA, and a cut-off energy of 600 eV. The total energy is converged to $10^{-6}$ eV and the atomic forces to 0.001 eV/\AA. In addition, we used the conjugate gradient algorithm (IBRION = 2) and the Gaussian smearing method (ISMEAR = 0) with smearing of 0.05 eV. Spin polarization was included in all calculations (ISPIN = 2). Symmetry was treated with the VASP default ($\mathrm{ISYM}=2$).

For band structure calculations, we use the $k$-point path generated by SeeK-path \cite{hinuma2017band,togo2024spglib}. A list of the coordinates of the high symmetry $k$-points is available in the Supplementary Material \cite{supplemental}.

To obtain the defect formation energies, we use the supercell approach. Specifically, we consider the fully relaxed bulk primitive cell, transform it into a conventional unit cell and build $4\times4\times1$ pristine and defective supercells. Furthermore, we use the same settings as for the bulk structure, except the convergence of the atomic forces that is set to 0.01 eV/$\AA$\; and the $k$-points consist of a mesh of $3\times 3 \times 1$.

The supercell approach considers the total energies of the supercells obtained with DFT and the chemical potentials of the bulk elemental phases of In and Se, which correspond to the total energies of these phases. We take as references the hexagonal Se (P3$_{1}$21, 152) and the tetragonal In (I4/mmm, 139), using the same settings as for the bulk $\alpha$-In$_2$Se$_3$ and $k$-point meshes of $10\times10\times 9$ and $12\times12\times 9$ respectively.\\
All defects are treated in the neutral charge state.
To compute the formation energy $E_f$ of a defect we consider \cite{hashibon2011first, xue2013first}:
\begin{equation}\label{eqn:formation_energy}
    E_{f} = E_f^0 - \frac{1}{2}\Delta G \Delta n_{In} - \Big(\Delta n_{Se} - \frac{3}{2}\Delta n_{In} \Big) \Delta \mu_{Se}~,
\end{equation}
where $\Delta G$ is the Gibbs free energy in the process of formation of one formula unit of In$_2$Se$_3$, $\Delta \mu$ is the difference of the chemical potential $\mu$ with respect to its elemental reference $\mu^{0}$, and $\Delta n$ is the number of particles added (a positive integer) or removed (a negative integer)  from the pristine supercell. We note that since our calculations are performed with DFT at $T=0$~K, $\Delta G = \Delta H - T\Delta S$ reduces to $\Delta H$, the enthalpy of formation, evaluated from total energies and therefore excluding zero-point contributions. Furthermore, $E_f^0$ is given by:
\begin{equation}
    E_f^0 = E^\mathrm{defective} - E^\mathrm{pristine} - \Delta n_{In} \mu_{In}^{0} - \Delta n_{Se} \mu_{Se}^{0}~,
\end{equation}
where $E^\mathrm{defective}$ is the total energy of the defective supercell, $E^\mathrm{pristine}$ is the total energy of the pristine non defective supercell, and $\mu_{Se}^{0}$ ($\mu_{In}^{0}$ ) is the reference chemical potential of bulk Se (In).

The equations that account for the In and Se rich environments are detailed in the Supplemental Material \cite{supplemental}.

\section{Results and Discussion}
\subsection{Bulk $\alpha$-In$_2$Se$_3$-3R}\label{sec:bulk_In2Se3}

\begin{figure}[h!]
    \centering
\includegraphics[width=1\columnwidth]{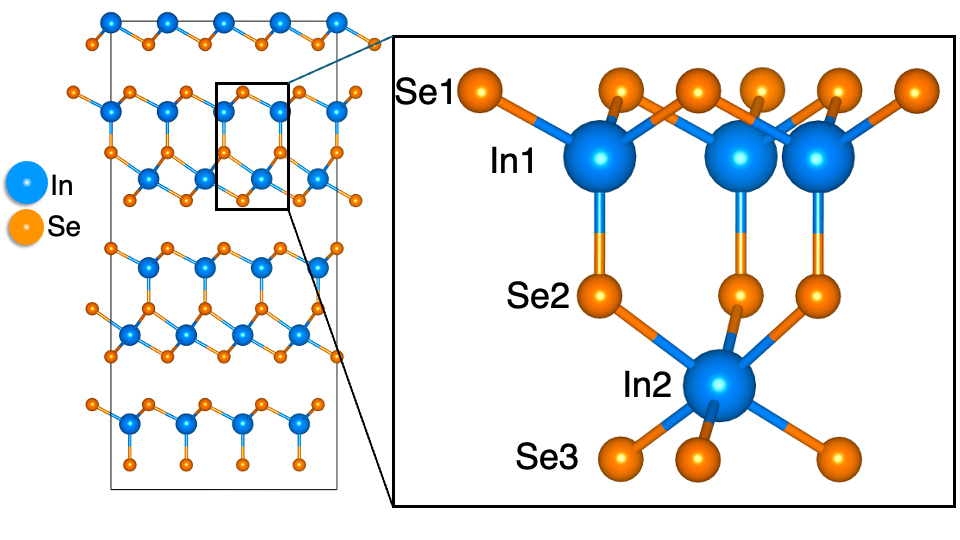}
 \caption{Atomic structure of $\alpha$-In$_2$Se$_3$-3R. On the left we show the bulk layered structure and on the right, one portion of one layer is magnified to show the tetrahedral and octahedral arrangements of In1 and In2 respectively.}  
 \label{fig:sketch_In2Se3}
\end{figure}

The bulk $\alpha$-In$_2$Se$_3$-3R is shown in Fig.~\ref{fig:sketch_In2Se3}. It has a layered structure consisting of Se1-In1-Se2-In2-Se3, i.e., three layers of Se and two layers of In, and belongs to the R3$m$ (160) space group. Our computed lattice constants are shown in Tab.~\ref{bulk_lattice_constants}. The lattice parameters $a$ and $b$ given by $r^2$SCAN+rVV10 are very close to experiment, whereas $c$ is slightly overestimated. Furthermore, we note the In1 atoms are connected tetrahedrally with the nearby Se atoms. Specifically, our relaxed structure consists of In1 atoms bonded to three Se(1) atoms separated at 2.626 \AA \, whereas Se(2) closes the tetrahedron vertex and lies perpendicularly at 2.536 \AA. On the other hand, In(2) is connected octahedrally to six neighboring Se. From our calculations, In(2) connects equidistantly to three Se(3) atoms at 2.691 \AA, and to three Se(2) atoms at 2.851 \AA. These predicted bond distances are close to the experimental estimations made in Ref.~\onlinecite{kupers2018controlled} with 2.55 to 2.64 \AA\, for the tetrahedral In1, and 2.71 to 2.85 \AA\, for the octahedral In(2).

With the elemental references mentioned in Sec.~\ref{sec:comp_methods}, we obtain $\Delta G = -3.36$~eV per formula unit, or $-323.9$~kJ\,mol$^{-1}$. This single quantity fixes the whole range over which the defect formation energies shift between the Se-rich and In-rich limits, see Eq.~\eqref{eqn:formation_energy} and the ones presented in the Supplementary Material \cite{supplemental}. The recommended value is given by Ref.~\onlinecite{chatillon1993} with $-275.8 \pm 25.0$~kJ\,mol$^{-1}$. Our value exceeds this recommendation in magnitude by $48.1$~kJ\,mol$^{-1}$ and thus falls outside its stated $\pm 25.0$~kJ\,mol$^{-1}$ uncertainty. Nevertheless, it lies within the spread of the individual experimental results collected in the same reference, which range from $-275.8$ to $-343.9$~kJ\,mol$^{-1}$, and it is consistent with the reported uncertainties of two of those measurements. We therefore conclude that r$^2$SCAN+rVV10 captures the thermochemistry of this compound within the experimental scatter, although not within the uncertainty of the recommended value.


\begin{table}[ht]
\centering
\caption{Lattice parameters and band gaps of bulk $\alpha$-In$_{2}$Se$_{3}$. The value of 1.25 eV was obtained by ARPES in Ref.~\onlinecite{kremer2025mexican}. We note that there is also an estimation of the fundamental band gap of 1.1 eV, as reported in Ref.~\onlinecite{li2018large}. The value in parenthesis of $E_g$ for r$^2$SCAN+rVV10 corresponds to the calculation performed with SOC.}
\begin{tabular}{c c c c}
\hline\hline  
Method                & $a=b$ (Å) & $c$ (Å) & $E_g$(eV) \\ 
\hline
r$^2$SCAN+rVV10      & 4.04    & 29.01  & 0.96(0.90)  \\
Experiment  & 4.03 \cite{kupers2018controlled}    & 28.75 \cite{kupers2018controlled} & 1.25 \cite{kremer2025mexican}  \\
\hline
\hline
\end{tabular}
\label{bulk_lattice_constants}
\end{table}

\begin{figure}[htb!]
    \centering
\includegraphics[width=1\columnwidth]{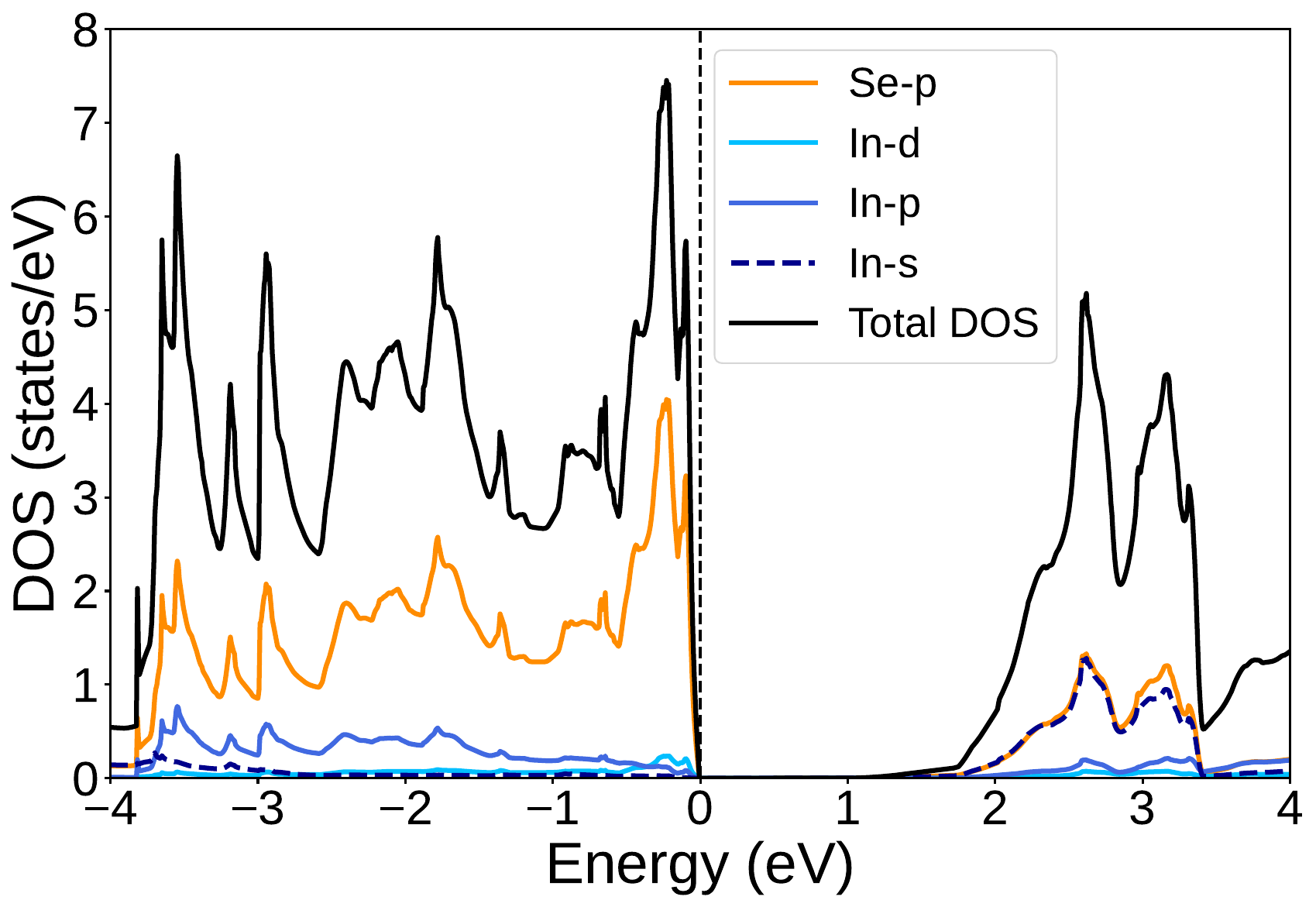}
 \caption{$\alpha$-In$_{2}$Se$_{3}$-3R DOS calculated with $r^2$SCAN+rVV10 and no SOC.}  
 \label{fig:dos_in2se3}
\end{figure}

\begin{figure}[htb!]
    \centering
\includegraphics[width=1\columnwidth]{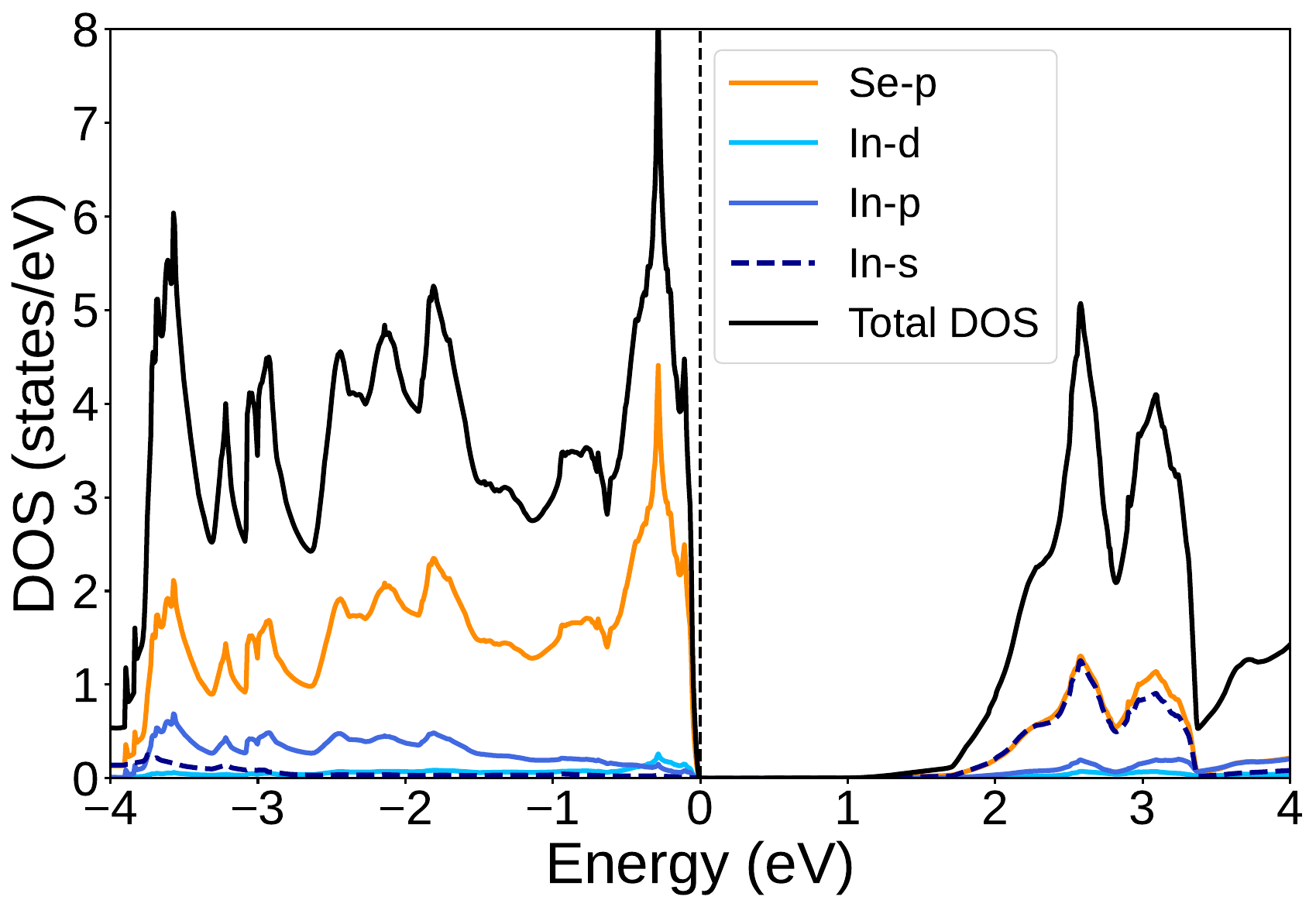}
 \caption{$\alpha$-In$_{2}$Se$_{3}$-3R DOS calculated with r$2$SCAN+rVV10 and SOC.}  
 \label{fig:dos_soc_in2se3}
\end{figure}

\begin{figure}[h!]
    \centering
\includegraphics[width=1\columnwidth]{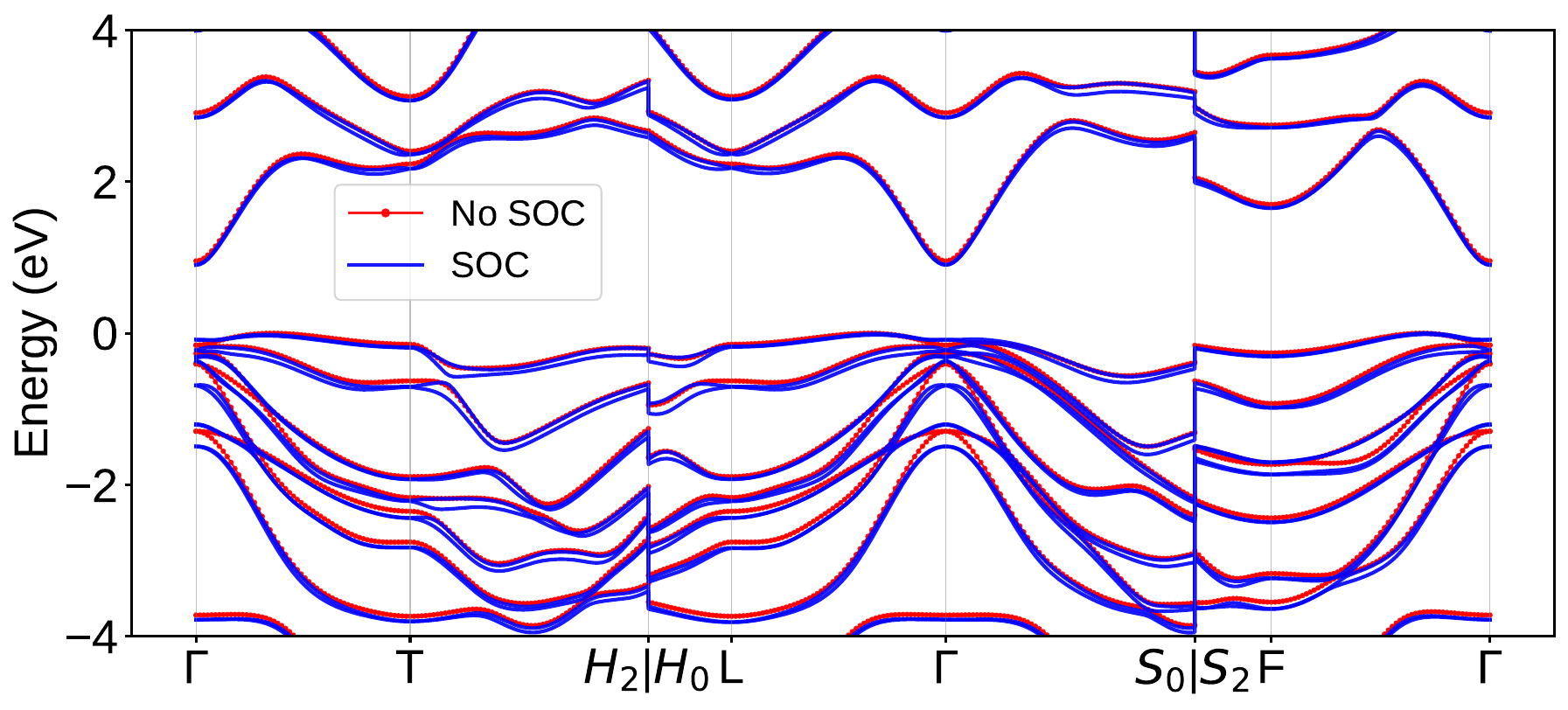}
 \caption{Band structure of $\alpha$-In$_{2}$Se$_{3}$-3R calculated with $r^2$SCAN+rVV10 with and without SOC. The bands are depicted in blue and red respectively and have been shifted with respect to the VBM.} 
 \label{fig:band_structure}
\end{figure}


\subsection{Electronic Structure}
Density of States (DOS) and band structure calculations with and without spin-orbit coupling (SOC) are performed using the primitive cell of $\alpha-$In$_2$Se$_3$-3R. Our results are shown in Figs.~\ref{fig:dos_in2se3}$-$\ref{fig:band_structure}.
The calculations without SOC depict an indirect bandgap of 0.96 eV. The valence band maximum (VBM) is located between F and $\Gamma$, and the conduction band minimum (CBM) occurs at the $\Gamma$ point. When SOC is considered, the band gap is reduced to 0.90 eV while preserving the indirect gap character. Furthermore, splitting of the bands close to the VBM is observed, whereas the CBM remains comparatively unaffected. The VBM is dominated by Se-p states, while the In-s and Se-p states prevail in the CBM. On the other hand, there are small contributions by the In-p and In-d states close to both the VBM and CBM.

The bandgap given by $r^2$SCAN+rVV10 underestimates the experimental value of 1.25 eV \cite{kremer2025mexican}. Calculations performed at the GGA(PBE) level provide a larger underestimation of the bandgap, whereas the hybrid (HSE) provide closer bandgaps to experiment \cite{li2018large, li2020band, kremer2025mexican}. Furthermore, our band structure results capture the so-called Mexican hat-like valence band dispersion, a particular feature observed in angle-resolved photoelectron spectroscopy (ARPES) measurements of $\alpha$-In$_2$Se$_3$-3R \cite{kremer2025mexican,mondal2026successful}. Importantly, this particular band-structure feature is linked to the strong DOS peak near the VBM seen in Fig.~\ref{fig:dos_in2se3} and further enhanced in our SOC calculations in Fig.~\ref{fig:dos_soc_in2se3}; this feature is known as a Van Hove singularity \cite{kremer2025mexican}. The latter has potential implications on magnetic and superconducting phase transitions \cite{cao2015tunable, kremer2025mexican, felton2025probing}.





\section{Defects in $\alpha$-$\mathrm{In_{2}Se_{3}}$}
In this section, we present the defect formation energies for a set of point defects in $\alpha$-In$_{2}$Se$_{3}$-3R. In this system, the positions of different In and Se atoms are not equivalent, hence a screening of vacancies and anti-sites in all layers is needed. Since there are three (two) Se (In) layers, we considered one vacancy and one anti-site per layer per point defect. Further, for understanding and guiding experimental synthesis, it is important to consider the formation energies at excess or deficiency of Se, i.e., Se- and In-rich conditions, respectively. The defect formation energies are shown in Fig.~\ref{in2se3_main}.

\subsection{Se-rich conditions}
Close inspection of Fig.~\ref{in2se3_main} shows that under Se-rich conditions, the antisite  ($Se_{In2}$) has the lowest defect formation energy with 1.24 eV, followed closely by the other antisite ($Se_{In1}$) and the Se vacancy ($V_{Se2}$), nearly degenerate with 1.44 eV and 1.45 eV, respectively. Upper in the energy scale we have a set of vacancies ($V_{Se1}$, $V_{Se3}$, and $V_{In1}$) with defect formation energies around 2.46-2.57 eV.  
The In vacancy ($V_{In2}$) follows with 3.12 eV and all the antisites $In_{Se}$ have the largest defect formation energies in the range 4.14 eV-4.66 eV, indicating they are unlikely to form under Se-rich conditions. Most of the defect formation energies match the expected behavior under an environment with excess of Se: antisites $Se_{In}$ will be more likely to occur than the $In_{Se}$ ones, and In vacancies will be easier to form. However, the lower energy of the Se vacancy ($V_{Se2}$) in the middle layer is intriguing: with more Se available, we would not expect this vacancy to have a low defect formation energy as the $Se_{In}$ antisites.
We rationalize the low formation energy of $V_{Se2}$ below.


\subsection{In-rich conditions}
\begin{figure}[h!]
    \centering
\includegraphics[width=1\columnwidth]
{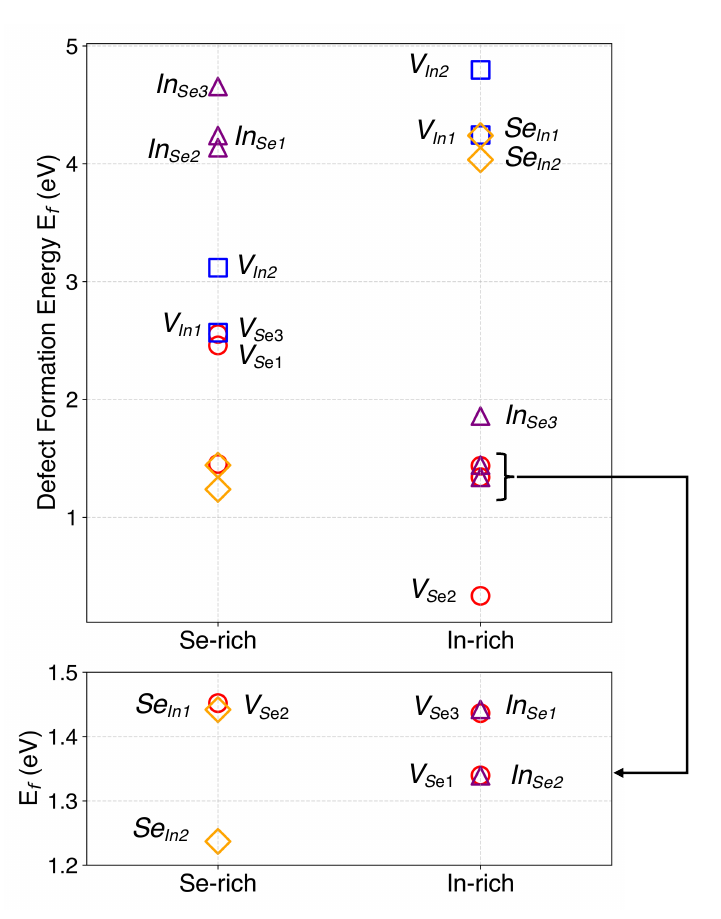}
 \caption{Defect formation energies of vacancies and antisite defects in $\alpha$-In$_2$Se$_3$. As shown in Fig.~\ref{fig:sketch_In2Se3}, we label the three layers of Se as 1 (top), 2 (middle) and 3 (bottom); and the two layers of In, labeled as 1 (top), and 2 (bottom). We show two panels: the top shows the defect formation energies of all the defects we studied and the bottom depicts a smaller energy range to provide a clearer view of the defect formation energies that are very close. The Se (In) vacancies are depicted through red (blue) circles (squares), and the In$_{Se}$ (Se$_{In}$) antisites are represented with purple (yellow) triangles (diamonds).}  
 \label{in2se3_main}
\end{figure}

Looking closely at the In-rich conditions, the vacancy of Se in the middle Se layer, $V_{Se2}$, is by far the most likely defect to occur under In-rich conditions with a formation energy of only 0.33 eV. Moreover, there is a small energetic region from around 1.3 eV to 1.44 eV (amplified in the right-hand side of the bottom panel of Fig.~\ref{in2se3_main}) where the vacancies $V_{Se1}$ and $V_{Se3}$, and the antisites $In_{Se1}$ and $In_{Se2}$ are competing. Located near this energetic region, the antisite $In_{Se3}$ appears with a formation energy of 1.86 eV. Furthermore, higher on the energy scale, from 4.03 eV to 4.25 eV, the antisites $Se_{In1}$ and $Se_{In2}$, and the vacancy $V_{In1}$ are competing and are unlikely to form. Finally, the $V_{In2}$ defect exhibits the largest formation energy, with 4.80 eV, showing it would be difficult to form In vacancies in the In bottom layer.

We rationalize the In and Se rich observations in the next subsection.

\subsection{Rationalizing the defect formation energies of $\alpha$-In$_2$Se$_3$-3R}
A remarkable observation of the Se-rich conditions is the Se vacancy $V_{Se2}$ competing with the $Se_{In}$ antisites, which is a trend one would generally not expect for a Se vacancy in a Se-rich environment. Nonetheless, a similar result was identified in Ref.~\onlinecite{ma2025vacancy} for a monolayer of $\alpha$-In$_2$Se$_3$, and in Ref.~\onlinecite{huang2021more} for Sb$_2$Se$_3$. Furthermore, Fig.~$1$(d) in Ref.~\onlinecite{zhou2017out} showed an intensity profile of one layer of $\alpha$-In$_2$Se$_3$: the Se layer in the middle ($Se2$) is the layer with the lowest intensity, followed by a higher intensity in the Se bottom layer ($Se3$) and an even higher intensity in the upper Se layer ($Se1$), suggesting the middle layer has more defects. In addition, Ref.~\onlinecite{zhou2017out} performed piezoresponse force microscopy (PFM) measurements and attributed the substantial leakage current to a high concentration of native defects, most likely Se vacancies, together with charge carriers.
Therefore, our findings for the selenium vacancy in the middle layer, $V_{Se2}$, are supported by both theoretical and experimental studies reported in the literature. 

The $V_{Se2}$ low defect formation energy ordering is counterintuitive on bond-counting grounds alone. As summarized in Tab.~\ref{tab:pristine_distances}, the $Se2$ atom is the only four-fold coordinated Se site in the structure: it bonds to one tetrahedral $In1$ at $2.536$~\AA\ along the three-fold axis, and to three octahedral $In2$ at $2.851$~\AA\ in the plane perpendicular to $c$, whereas $Se1$ and $Se3$ are three-fold coordinated at $2.626$~\AA\ and $2.691$~\AA, respectively. Forming $V_{Se2}$ therefore severs more In--Se bonds than either alternative but still costs the least energy. In the following, we discuss two arguments that support $V_{Se2}$ having a particularly low defect formation energy: the markedly unequal strengths of the bonds that are broken, and the site symmetry that survives their removal.

\subsubsection{Bond strengths}

Following the bond-valence model~\cite{brown1985bond, brese1991bond}, we assign to each In--Se bond a valence $s=\exp[(R_{0}-d)/b]$ with $b=0.37$~\AA\ and
$R_{0}=2.47$~\AA, so that short bonds are strong and long bonds are weak, as
expected on simple chemical grounds. At fixed $b$, changing $R_{0}$ multiplies every In--Se valence by the common factor $\exp(\Delta R_{0}/b)$, so every ratio of valence sums quoted below is independent of $R_{0}$, whereas the absolute sums are not. The sum $\sum s$ over the bonds broken by a
vacancy is then a rough measure of the bonding that must be destroyed to create
it. The resulting values are collected in Tab.~\ref{tab:bond_valence}.

The four bonds at the $Se2$ site are strikingly unequal. The axial $In1$--$Se2$ bond is the shortest in the structure and by far the strongest ($s=0.837$~v.u.), while the three basal $In2$--$Se2$ bonds are the longest and weakest ($s=0.357$~v.u.~each). Their sum, $\sum s = 1.909$~v.u., is smaller than the $1.968$~v.u.~accumulated by the three equivalent, considerably stronger bonds at the $Se1$ site, even though one more bond is cut. Removing $Se2$ thus destroys one strong bond and three weak ones, and in total slightly less bonding than removing $Se1$. To this extent, the bond-strength picture is consistent with the ordering $E_{f}(V_{Se2}) < E_{f}(V_{Se1})$, although a $3\%$ difference in $\sum s$ should not be read as accounting for its magnitude.

The bond hierarchy inferred here from geometry alone is corroborated by an independent electronic structure analysis. Reference \onlinecite{kupers2018controlled} evaluated crystal orbital Hamilton populations (COHP) for 3R-In$_2$Se$_3$ and found the tetrahedral $In1$ environment to be stabilized by covalent In--Se interactions, in contrast to the more ionic octahedral $In2$ environment. Three of the four bonds at $Se2$ are to octahedral $In2$ centers, so the bonding at that site is dominated by the weaker and less covalent part of the network, which is the same conclusion reached from the bond valences. 

However, the bond-valence model alone does not explain why $E_{f}(V_{Se2}) < E_{f}(V_{Se3})$. The $Se3$ site carries the weakest bonding of the three, with $\sum s = 1.650$~v.u., around $16\%$ below $Se1$, yet $V_{Se3}$ is the most expensive of the three Se vacancies. The correlation between broken bond valence and formation energy is therefore not merely weak but inverted for $Se3$. This can be understood by considering the symmetry after formation of the defect and the energy gain from the relaxation of the defective structure, as we describe next.

\begin{table}[tb]
\caption{\label{tab:bond_valence}%
Bond valences at the Se sites of $\alpha$-In$_2$Se$_3$-3R and Bi$_2$Se$_3$,
computed from the relaxed r$^2$SCAN+rVV10 structures. $d$ is the cation--Se
distance and $s=\exp[(R_{0}-d)/b]$
the bond valence in valence units (v.u.), with $b=0.37$~\AA,
$R_{0}(\text{In--Se})=2.47$~\AA\, and $R_{0}(\text{Bi--Se})=2.72$~\AA\
taken from Ref.~\onlinecite{brese1991bond}. The sum $\sum s$ runs over all bonds at the
site and measures the total bonding that a vacancy on that site must destroy.
Because $R_{0}$ differs between In--Se and Bi--Se, only ratios formed within a
single compound are independent of that empirical parameter. The last column
is therefore the ratio of $\sum s$ to its value at the outer Se site of the
same compound ($Se1$ for $\alpha$-In$_2$Se$_3$, the outer Se of the quintuple
layer for Bi$_2$Se$_3$).
Sums are evaluated from the full-precision relaxed distances, so the last digit of $\sum s$ may differ by one unit from the product of the rounded entries shown. A detailed derivation, together with the distances at full precision, is given in the Supplemental Material~\cite{supplemental}.}
\begin{ruledtabular}
\begin{tabular}{llcccc}
Site & Bond & $d$ (\AA) & $s$ (v.u.) & $\sum s$ (v.u.) & Ratio \\
\colrule
\multicolumn{6}{l}{$\alpha$-In$_2$Se$_3$-3R ($R3m$)}\\
$Se1$ & $3\times In1$              & 2.626 & 0.656 & 1.968 & 1.00 \\
$Se2$ & $1\times In1$\footnotemark[1] & 2.536 & 0.837 & 1.909 & 0.97 \\
      & $3\times In2$\footnotemark[2] & 2.851 & 0.357 &       &      \\
$Se3$ & $3\times In2$              & 2.691 & 0.550 & 1.650 & 0.84 \\
\colrule
\multicolumn{6}{l}{Bi$_2$Se$_3$ ($R\bar{3}m$)}\\
Se (outer)   & $3\times$Bi & 2.856 & 0.692 & 2.077 & 1.00 \\
Se (central) & $6\times$Bi & 3.079 & 0.379 & 2.273 & 1.09 \\
\end{tabular}
\footnotetext[1]{Axial neighbor, along the three-fold axis.}
\footnotetext[2]{Basal neighbors, in the plane perpendicular to $c$.}
\end{ruledtabular}
\end{table}

\subsubsection{Symmetry of the relaxation}

Before turning to the relaxation, it is worth noting what the coordination polyhedra alone can and cannot explain. $V_{Se1}$ affects only tetrahedra (three $In1$ centers whose coordination changes as $4 \rightarrow 3$) and $V_{Se3}$ only octahedra (three $In2$ centers whose coordination changes as $6 \rightarrow 5$), whereas $V_{Se2}$ is the only vacancy that dismantles both structures at once: it reduces the coordination of one tetrahedral $In1$ as $4 \rightarrow 3$ and the coordination of three octahedral $In2$ as $6 \rightarrow 5$. Still, $V_{Se2}$ is the cheapest of the three defects to form. Regardless of which coordination In is taken to prefer, the polyhedral argument cannot predict the observed energetic ordering of these defects, since the vacancy that destroys the most structure is the one with the lowest formation energy.

What distinguishes $V_{Se2}$ is instead the symmetry that survives. In the polar space group $R3m$ the tetrahedral $In1$ center has local $C_{3v}$ symmetry, with $Se2$ as its unique axial neighbor and the three $Se1$ atoms as its basal neighbors. Every atom in this structure lies on a three-fold axis, so all three Se vacancies retain $C_{3v}$ point symmetry as defects. What singles out $Se2$ is that the axis through the vacant site also passes through the tetrahedral $In1$ that loses a ligand, so that this $In1$ alone keeps its $C_{3v}$ site symmetry; the three octahedral $In2$ it also affects lie off that axis, as do the affected cations of $V_{Se1}$ and $V_{Se3}$.

The three-fold axis runs along $c$ through $In1$ and the $Se2$ site, so $Se2$ is the only Se in the structure whose removal leaves that axis intact. 

\begin{figure}[h!]
    \centering
\includegraphics[width=1\columnwidth]
{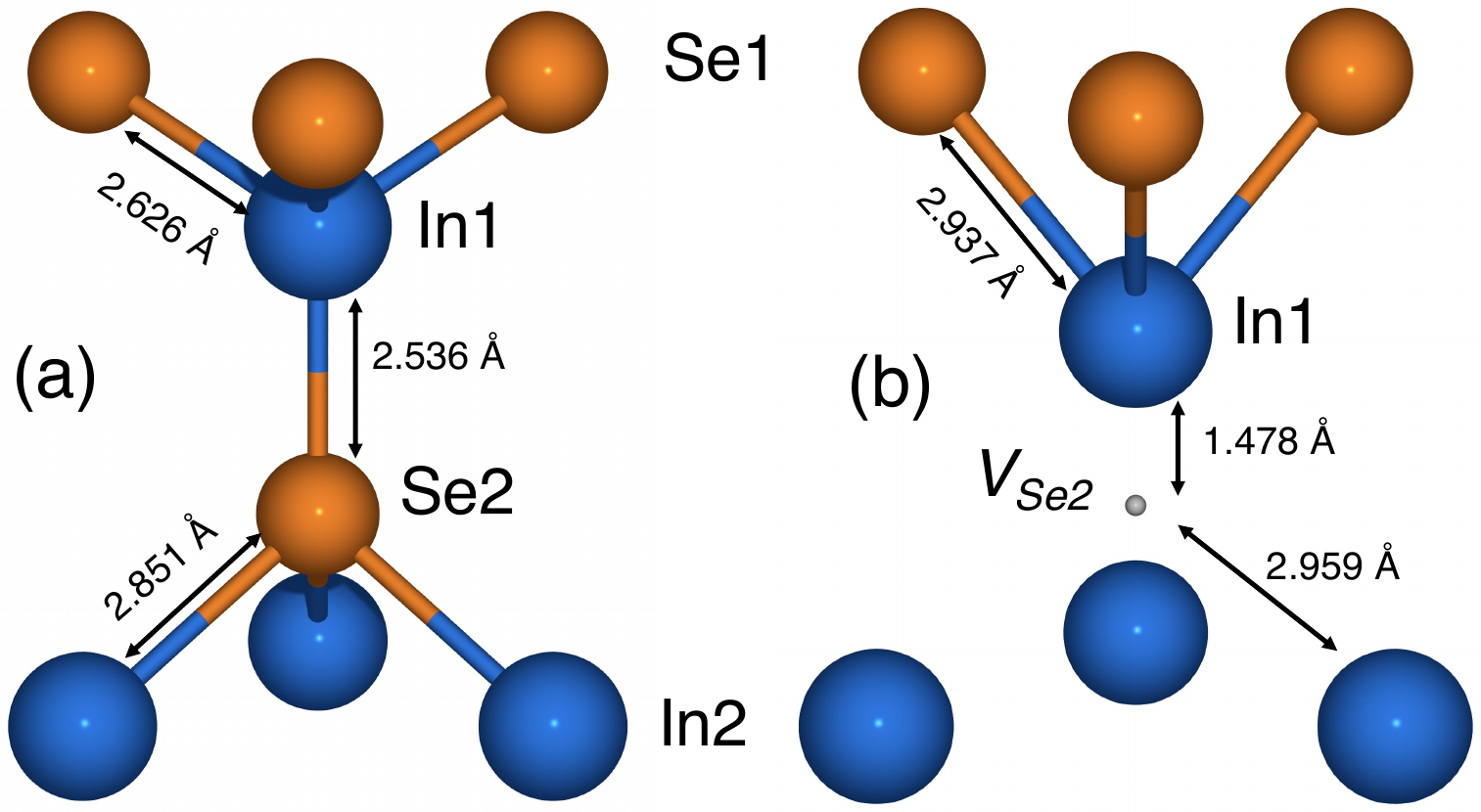}
 \caption{(a) Sketch of the pristine $\alpha$-In$_2$Se$_3$. (b) Sketch of the vacancy of Se ($V_{Se2}$) in the middle layer. A tiny gray dot represents the site of the original Se. In both frames we do not show the Se(3) layer for clarity.}  
 \label{fig:Sketch_V_Se2}
\end{figure}

The consequences are evident in Tab.~\ref{tab:defect_distances} and Fig.~\ref{fig:Sketch_V_Se2}. Upon removal of $Se2$, the neighboring $In1$ relaxes along the surviving axis, moving $1.058$~\AA\ towards the vacant site without lowering its site symmetry. At the same time, that $In1$'s three $In1$--$Se1$ bonds elongate uniformly from $2.626$ to $2.937$~\AA\ and the three corresponding $Se1$--$In1$--$Se1$ angles contract uniformly from $100.56^{\circ}$ to $76.06^{\circ}$. $In1$ is left three-fold coordinated in a strongly pyramidal geometry pointing towards the vacancy. The three octahedral $In2$ neighbors, each of which loses only one of its six ligands, are barely perturbed and displace by just $0.144$~\AA. The relaxation is thus large, unfrustrated, and concentrated on the single cation that is free to move.

In contrast, for $V_{Se1}$ and $V_{Se3}$ the vacant Se is instead one of three equivalent neighbors of each affected cation, so any relaxation must break the local three-fold axis. The response is correspondingly frustrated and small: for $V_{Se1}$ the three $In1$ centers displace by only $0.382$~\AA, almost entirely perpendicular to $c$ ($0.380$~\AA\ in that plane against $0.036$~\AA\ along $c$). In addition, the surviving bonds split into $2.564$ and $2.682$~\AA\ (a spread of $0.118$~\AA) with an angular spread of $17.71^{\circ}$. For $V_{Se3}$ the three octahedral $In2$ displace $0.440$~\AA\ away from the vacant site ($0.301$~\AA\ perpendicular to $c$ and $0.321$~\AA\ along it), so that their separation from the vacant site grows from $2.691$ to $3.115$~\AA\ and the $In2 \cdots In2$ distance across the vacancy expands from $4.040$ to $4.561$~\AA. In both cases, the surviving coordination shell is nonequivalent. As a consequence, the formation energy is more than $1$~eV higher than that of $V_{Se2}$. Further displacements and angles are tabulated in the Supplemental Material~\cite{supplemental}.

The low formation energy of $V_{Se2}$ therefore arises from the interplay of two effects: three of its four bonds are the weakest In--Se bonds in the structure and, although it is the only vacancy to break both coordination polyhedra, it is also the only one that leaves the affected tetrahedral $In1$ on the axis through the vacant site, so that the lattice can respond with a large, unfrustrated, symmetry-adapted relaxation rather than a frustrated one. This relaxation does not, however, restore the bonding that was lost. Applying the same bond-valence parameters to the relaxed structure of Tab.~\ref{tab:defect_distances}, the valence sum at the relaxing $In1$ falls from $2.805$ to $0.849$~v.u., and even the neighboring $Se1$ and $In2$ centers end slightly below their pristine values ($1.968 \rightarrow 1.822$ and $2.722 \rightarrow 2.646$~v.u.) despite the contraction of their surviving bonds. The bond-valence model therefore measures the bonding that a vacancy destroys, but not the stabilization that follows the relaxation. 

\subsubsection{Energy gain by relaxation of the defective structure}

The importance of the relaxation can be assessed directly by comparing the defect formation energy obtained at the relaxed structure, $E_{f}^{\mathrm{relaxed}}$, with that of the unrelaxed structure,  $E_{f}^{\mathrm{unrelaxed}}$, i.e., when keeping the lattice constant and atomic positions fixed when creating the defective system. The difference $E_{f}^{\mathrm{relaxation}} = E_{f}^{\mathrm{unrelaxed}} - E_{f}^{\mathrm{relaxed}}$ is the
energy gained by relaxation of the defective structure. For the unrelaxed structures, we find that $V_{Se2}$ and $V_{Se3}$ are degenerate to within $7$~meV, and $V_{Se1}$ is $0.299$~eV above $V_{Se2}$. However, the energy gain by relaxation between $V_{Se2}$ and $V_{Se3}$ differs by more than a factor of six: We find $1.293$~eV for $V_{Se2}$, the largest gain of all defects considered in this work, $0.586$ for $V_{Se1}$, and $0.197$~eV for $V_{Se3}$. Within these three, the ordering of the formation energies is therefore set by the energy gain through relaxation and not by the unrelaxed lattice: $V_{Se3}$ relaxes the least and is the most expensive Se vacancy, while $V_{Se2}$ relaxes the most and is the cheapest vacancy to form. This establishes that the relaxation is decisive and thereby complements the analysis of the symmetry change under relaxation.
See the Supplemental Material for relaxation energies of all defects considered in Fig.~\ref{in2se3_main} \cite{supplemental}.


\subsubsection{Comparison with the centrosymmetric structure type}
It is instructive to ask how much of this argument depends on the polar structure, and what the analysis above would give for a centrosymmetric one such as $\beta$-In$_2$Se$_3$, the phase into which $\alpha$-In$_2$Se$_3$ transforms upon heating~\cite{miyazawa1957}, and which adopts the same tetradymite structure as the prototypical Bi$_2$Se$_3$: space group $R\bar{3}m$, quintuple layers, cation sites that are exclusively octahedral, two equivalent outer Se layers, and a central Se layer lying on an inversion center with six equivalent cation neighbors. The axial/basal distinction at $In1$, the inequivalence of the two outer Se layers, and the unique axial neighbor of $Se2$ are thereby removed altogether.

Note that in the centrosymmetric $\beta$-structure the central Se of each quintuple layer binds three cations on either side, whereas in the $\alpha$-structure it binds one $In1$ axially and three $In2$ basally. That asymmetry is what removes the inversion center and thus allows for the spontaneous polarization for which $\alpha$-In$_2$Se$_3$ is often studied~\cite{ding2017prediction, zhou2017out}. The same asymmetry also supplies the two ingredients identified above: one short axial bond against three long basal ones, and a three-fold axis running through $In1$ and the $Se2$ site.

To the best of our knowledge, no defect formation energies have been reported for $\beta$-In$_2$Se$_3$, and we leave them for future work. Instead, we consider Bi$_2$Se$_3$, for which native-defect energies are readily available~\cite{west2012}. There, the lowest-energy Se vacancy is reported to be the one in the outer layer, an ordering that Ref.~\onlinecite{west2012} attributes to bond counting alone, an anion carrying six bonds at the center of a quintuple layer being more stable than one carrying three at its surface. Our bond-valence analysis of pristine Bi$_2$Se$_3$, relaxed with the same r$^2$SCAN+rVV10 setup, comes to the same conclusion: its central Se carries $1.09$ times the bonding of its outer Se, whereas in $\alpha$-In$_2$Se$_3$ the corresponding ratio is $0.97$ (Tab.~\ref{tab:bond_valence}). In Bi$_2$Se$_3$ the cheapest Se vacancy is thus the site with the fewest bonds, while in $\alpha$-In$_2$Se$_3$ it is the site with the most. Simple bond counting accounts for the centrosymmetric structure and fails for the polar one. Further details are given in the Supplemental Material~\cite{supplemental}.


\begin{table}[h!]
\centering
\caption{Nearest-neighbor distances in pristine $\alpha$-In$_2$Se$_3$-3R, in \AA, with the multiplicity in parentheses. A dash means the two sites are not nearest neighbors. $N_c$ is the coordination number, equal to the sum of the multiplicities in the row. Only adjacent layers are bonded, so each site couples to at most two others.}
\begin{tabular}{cccccccc}
\hline\hline
 & Se1 & In1 & Se2 & In2 & Se3 & $N_c$ \\
\hline
Se1 & --        & 2.626~(3) & --        & --        & --        & 3 \\
In1 & 2.626~(3) & --        & 2.536~(1) & --        & --        & 4 \\
Se2 & --        & 2.536~(1) & --        & 2.851~(3) & --        & 4 \\
In2 & --        & --        & 2.851~(3) & --        & 2.691~(3) & 6 \\
Se3 & --        & --        & --        & 2.691~(3) & --        & 3 \\
\hline\hline
\end{tabular}
\label{tab:pristine_distances}
\end{table}

\begin{table}[h!]
\centering
\caption{Distances after forming $V_{Se2}$, in \AA, for the atoms adjacent to the vacancy, with the multiplicity in parentheses. The $V_{Se2}$ row and column give distances to the vacant site, i.e., to the position that the removed Se2 occupied, and are therefore separations, not bonds; they are excluded from the coordination count $N_c$. The corresponding values of the pristine cell are 2.536~\AA\ to In1 and 2.851~\AA\ to In2. Because the vacancy makes atoms at the same pristine site inequivalent, each row describes a specific atom, identified in the footnotes, rather than a site type. As a consequence, the resulting matrix is not symmetric. In1$^{b}$ binds three equivalent Se1 at 2.937~\AA, whereas each Se1$^{a}$ sees that atom only once while its other two In1 bonds are contracting to 2.567~\AA\ from the pristine 2.626~\AA\ as In1$^{b}$ withdraws. Far from the vacancy, the pristine values are recovered: an In1 more than 8~\AA\ away has Se1 at 2.627--2.630~\AA\ and Se2 at 2.535~\AA, against 2.626 and 2.536~\AA\ in the pristine cell.}
\begin{tabular}{ccccccc}
\hline\hline
 & Se1 & In1 & $V_{Se2}$ & In2 & Se3 & $N_c$ \\
\hline
Se1$^{a}$ & --        & 2.567~(2) & --      & --        & --        & 3 \\
          &           & 2.937~(1) &         &           &           &   \\
In1$^{b}$ & 2.937~(3) & --        & [1.478] & --        & --        & 3 \\
$V_{Se2}$ & --        & [1.478]   & --      & [2.959]   & --        & -- \\
In2$^{c}$ & --        & --        & [2.959] & --        & 2.631~(1) & 5$^{e}$ \\
          &           &           &         &           & 2.664~(2) &   \\
Se3$^{d}$ & --        & --        & --      & 2.664~(2) & --        & 3 \\
          &           &           &         & 2.714~(1) &           &   \\
\hline\hline
\end{tabular}

\begin{flushleft}
\footnotesize
$^{a}$ One of the three Se1 bonded to In1$^{b}$; its other two In1 neighbors lie
further from the vacancy and retain their axial Se2.\\
$^{b}$ The In1 that lost its axial Se2, the atom that relaxes 1.058~\AA\ along
$c$ into the vacancy.\\
$^{c}$ One of the three In2 that lost a Se2.\\
$^{d}$ A Se3 bonded to In2$^{c}$.\\
$^{e}$ In2$^{c}$ also retains two surviving Se2 neighbors at 2.802~\AA, counted in
$N_c$ but not listed, since the Se2 column here denotes the vacant site.
\end{flushleft}
\label{tab:defect_distances}
\end{table}



\section{Conclusion}
In this work, we investigated the structural, electronic, and defect properties of bulk $\alpha$-In$_2$Se$_3$-3R using density functional theory with the meta-GGA r$^2$SCAN+rVV10 exchange-correlation functional. The calculated lattice parameters, local bonding geometry, and characteristic features of the electronic structure, including the Mexican hat-like valence band dispersion, are in good agreement with available experimental measurements, demonstrating that r$^2$SCAN+rVV10 provides an accurate description of the structural and electronic properties of this layered material. Although the calculated band gap remains underestimated with respect to experiment, it represents an improvement over conventional GGA calculations while preserving the essential features of the band structure.

Our defect calculations reveal that the Se vacancy in the middle layer ($V_{\mathrm{Se2}}$) is one of the most thermodynamically favorable native defects under both Se-rich and In-rich growth conditions, competing only with a small number of antisite defects depending on the chemical environment. This behavior is consistent with previous theoretical studies and with experimental observations reporting a high concentration of Se vacancies in $\alpha$-In$_2$Se$_3$. Our analysis of the relaxed defective structure indicates that the remarkable stability of $V_{Se2}$ can be explained by the interplay of the markedly unequal strengths of the In--Se bonds in $\alpha$-In$_2$Se$_3$-3R and the site symmetry that survives their removal. The latter is associated with a local geometric relaxation, in which the neighboring tetrahedrally coordinated In atom displaces significantly toward the vacancy along the three-fold axis. This displacement preserves the $C_{3v}$ site symmetry of that In atom, while its coordination is reduced from four to three and its three surviving ligands adopt a strongly pyramidal arrangement.


These findings suggest that the specific symmetry of $\alpha$-In$_2$Se$_3$-3R and its structural relaxation after forming the $V_{Se2}$ vacancy play a central role in stabilizing this native defect.

Overall, the present work demonstrates that r$^2$SCAN+rVV10 constitutes a robust and computationally efficient framework for investigating native defects in layered chalcogenides. The methodology and physical insights presented here provide a foundation for future studies of charged defects, defect-induced electronic states, polarization--defect coupling, and defect engineering in ferroelectric two-dimensional materials.


\begin{acknowledgments}
This work was supported by the U.S. DOE, Office of Science, Basic Energy Sciences (BES), Grant No. DE-SC0014208. Z.Q.M. acknowledges the support from the US DOE under grants
DE-SC0019068. This research used resources of the National Energy Research
Scientific Computing Center, a DOE Office of Science User Facility
supported by the Office of Science of the U.S. Department of Energy
under Contract No. DE-AC02-05CH11231 using NERSC award
BES-ERCAP0036501. T.L. acknowledges support from the Alexander von Humboldt Foundation through a Feodor Lynen Research Fellowship.

\end{acknowledgments}

\appendix



\bibliography{bibliography}

@article{chatillon1993,
  author  = {Chatillon, Christian},
  title   = {Critical analysis of the thermodynamic properties
             of the {In--Se} gaseous and solid phases},
  journal = {J. Cryst. Growth},
  volume  = {129},
  pages   = {297--311},
  year    = {1993},
  doi     = {10.1016/0022-0248(93)90461-5}
}

@article{west2012,
  author  = {West, D. and Sun, Y. Y. and Wang, Han and
             Bang, Junhyeok and Zhang, S. B.},
  title   = {Native Defects in Second-Generation Topological
             Insulators: Effect of Spin-Orbit Interaction on
             {Bi$_2$Se$_3$}},
  journal = {Physical Review B},
  volume  = {86},
  pages   = {121201},
  year    = {2012},
  doi     = {10.1103/PhysRevB.86.121201}
}

@article{miyazawa1957,
  author  = {Miyazawa, Hisao and Sugaike, Suezo},
  title   = {Phase Transition of {In$_2$Se$_3$}},
  journal = {Journal of the Physical Society of Japan},
  volume  = {12},
  number  = {3},
  pages   = {312},
  year    = {1957},
  doi     = {10.1143/JPSJ.12.312}
}

@article{brese1991bond,
  title={Bond-valence parameters for solids},
  author={Brese, NE and O'keeffe, MJSS},
  journal={Structural Science},
  volume={47},
  number={2},
  pages={192--197},
  year={1991},
  publisher={International Union of Crystallography}
}

@article{brown1985bond,
  title={Bond-valence parameters obtained from a systematic analysis of the inorganic crystal structure database},
  author={Brown, I David and Altermatt, DJSS},
  journal={Structural Science},
  volume={41},
  number={4},
  pages={244--247},
  year={1985},
  publisher={International Union of Crystallography}
}

@article{hashibon2011first,
  title={First-principles density functional theory study of native point defects in Bi 2 Te 3},
  author={Hashibon, Adham and Els{\"a}sser, Christian},
  journal={Physical Review B—Condensed Matter and Materials Physics},
  volume={84},
  number={14},
  pages={144117},
  year={2011},
  publisher={APS}
}

@article{freysoldt2014first,
  title={First-principles calculations for point defects in solids},
  author={Freysoldt, Christoph and Grabowski, Blazej and Hickel, Tilmann and Neugebauer, J{\"o}rg and Kresse, Georg and Janotti, Anderson and Van de Walle, Chris G},
  journal={Reviews of modern physics},
  volume={86},
  number={1},
  pages={253--305},
  year={2014},
  publisher={APS}
}

@article{cao2015tunable,
  title={Tunable magnetism and half-metallicity in hole-doped monolayer GaSe},
  author={Cao, Ting and Li, Zhenglu and Louie, Steven G},
  journal={Physical review letters},
  volume={114},
  number={23},
  pages={236602},
  year={2015},
  publisher={APS}
}

@article{felton2025probing,
  title={Probing and manipulating the Mexican hat-shaped valence band of In2Se3},
  author={Felton, James and Harknett, Jordan and Page, Joe and Yang, Zhuo and Alghofaili, Nada and O’Shea, James N and Eaves, Laurence and Kohama, Yoshimitsu and Greenaway, Mark T and Patan{\`e}, Amalia},
  journal={Nature Communications},
  volume={16},
  number={1},
  pages={922},
  year={2025},
  publisher={Nature Publishing Group UK London}
}

@article{kremer2025mexican,
  title={Mexican hat--like valence band dispersion and quantum confinement in rhombohedral ferroelectric $\alpha$-In 2 Se 3},
  author={Kremer, Geoffroy and Mahmoudi, Aymen and Bouaziz, Meryem and Rahimi, Mehrdad and Bertran, Fran{\c{c}}ois and Dayen, Jean-Francois and Della Rocca, Maria Luisa and Pala, Marco and Naitabdi, Ahmed and Chaste, Julien and others},
  journal={Physical Review B},
  volume={112},
  number={15},
  pages={155418},
  year={2025},
  publisher={APS}
}

@article{togo2024spglib,
  title={Spglib: a software library for crystal symmetry search},
  author={Togo, Atsushi and Shinohara, Kohei and Tanaka, Isao},
  journal={Science and Technology of Advanced Materials: Methods},
  volume={4},
  number={1},
  pages={2384822},
  year={2024},
  publisher={Taylor \& Francis}
}

@article{hinuma2017band,
  title={Band structure diagram paths based on crystallography},
  author={Hinuma, Yoyo and Pizzi, Giovanni and Kumagai, Yu and Oba, Fumiyasu and Tanaka, Isao},
  journal={Computational Materials Science},
  volume={128},
  pages={140--184},
  year={2017},
  publisher={Elsevier}
}

@article{li2018large,
  title={Large disparity between optical and fundamental band gaps in layered In 2 Se 3},
  author={Li, Wei and Sabino, Fernando P and Crasto de Lima, Felipe and Wang, Tianshi and Miwa, Roberto H and Janotti, Anderson},
  journal={Physical Review B},
  volume={98},
  number={16},
  pages={165134},
  year={2018},
  publisher={APS}
}

@misc{supplemental,
note = {See Supplemental Material at [URL will be inserted by publisher] for the high-symmetry $k$-point path used in the band-structure calculations, the derivation of the defect formation energies in the Se-rich and In-rich limits, tabulated defect formation energies, the bond lengths and angles around each vacancy, and the bond-valence analysis of the Se sites.}
}

@article{ning2022workhorse,
  title={Workhorse minimally empirical dispersion-corrected density functional with tests for weakly bound systems: r 2 SCAN+ rVV 10},
  author={Ning, Jinliang and Kothakonda, Manish and Furness, James W and Kaplan, Aaron D and Ehlert, Sebastian and Brandenburg, Jan Gerit and Perdew, John P and Sun, Jianwei},
  journal={Physical Review B},
  volume={106},
  number={7},
  pages={075422},
  year={2022},
  publisher={APS}
}

@article{peng2016versatile,
  title={Versatile van der Waals density functional based on a meta-generalized gradient approximation},
  author={Peng, Haowei and Yang, Zeng-Hui and Perdew, John P and Sun, Jianwei},
  journal={Physical Review X},
  volume={6},
  number={4},
  pages={041005},
  year={2016},
  publisher={APS}
}

@article{perdew1996generalized,
  title={Generalized gradient approximation made simple},
  author={Perdew, John P and Burke, Kieron and Ernzerhof, Matthias},
  journal={Physical review letters},
  volume={77},
  number={18},
  pages={3865},
  year={1996},
  publisher={APS}
}

@article{sun2016accurate,
  title={Accurate first-principles structures and energies of diversely bonded systems from an efficient density functional},
  author={Sun, Jianwei and Remsing, Richard C and Zhang, Yubo and Sun, Zhaoru and Ruzsinszky, Adrienn and Peng, Haowei and Yang, Zenghui and Paul, Arpita and Waghmare, Umesh and Wu, Xifan and others},
  journal={Nature chemistry},
  volume={8},
  number={9},
  pages={831--836},
  year={2016},
  publisher={Nature Publishing Group UK London}
}

@article{scan,
  title={Strongly constrained and appropriately normed semilocal density functional},
  author={Sun, Jianwei and Ruzsinszky, Adrienn and Perdew, John P},
  journal={Physical review letters},
  volume={115},
  number={3},
  pages={036402},
  year={2015},
  publisher={APS}
}

@article{mondal2026successful,
  title={Successful growth of low carrier density $\alpha$-In$_2$Se$_3$ single crystals using Se-flux in a modified Bridgman furnace},
  author={Mondal, Soumi and Anil, Sreekant and Islam, Saurav and Guan, Yingdong and Ayyagari, Sai Venkata Gayathri and Pearre, Aaron and Santhosh, Sandra and Alem, Nasim and Samarth, Nitin and Mao, Zhiqiang},
  journal={arXiv preprint arXiv:2601.12588},
  year={2026}
}

@article{hou2020neutron,
  title={A neutron irradiation-induced displacement damage of indium vacancies in $\alpha$-In 2 Se 3 nanoflakes},
  author={Hou, Pengfei and Wang, Xinhao and Liu, Yunxia and Chen, Yun and Dong, Shijian and Guo, Hongxia and Wang, Jinbin and Zhong, Xiangli and Ouyang, Xiaoping},
  journal={Physical Chemistry Chemical Physics},
  volume={22},
  number={28},
  pages={15799--15804},
  year={2020},
  publisher={Royal Society of Chemistry}
}

@article{tang2021controllable,
  title={Controllable polarization and doping in ferroelectric In2Se3 monolayers and heterobilayers via intrinsic defect engineering},
  author={Tang, Cheng and Zhang, Lei and Wijethunge, Dimuthu and Ostrikov, Kostya Ken and Du, Aijun},
  journal={The Journal of Physical Chemistry C},
  volume={125},
  number={44},
  pages={24648--24654},
  year={2021},
  publisher={ACS Publications}
}

@article{xiao2018intrinsic,
  title={Intrinsic two-dimensional ferroelectricity with dipole locking},
  author={Xiao, Jun and Zhu, Hanyu and Wang, Ying and Feng, Wei and Hu, Yunxia and Dasgupta, Arvind and Han, Yimo and Wang, Yuan and Muller, David A and Martin, Lane W and others},
  journal={Physical review letters},
  volume={120},
  number={22},
  pages={227601},
  year={2018},
  publisher={APS}
}

@article{ma2025vacancy,
  title={Vacancy defects in 2D ferroelectric In2Se3 and the conductivity modulation by polarization--defect coupling},
  author={Ma, Ming-Yu and Wang, Dan and Huang, Yu-Ting and Han, Dong and Chen, Nian-Ke and Sun, Hong-Bo and Zhang, Shengbai and Li, Xian-Bin},
  journal={Nano Letters},
  volume={25},
  number={10},
  pages={3726--3732},
  year={2025},
  publisher={ACS Publications}
}

@article{pang2025scanning,
  title={Scanning Tunneling Microscopy Characterization of Intrinsic Point Defects and Their Local Density of States in $\alpha$-In2Se3},
  author={Pang, Wenhui and Liu, Zi and Li, Jieying and Cai, Desheng and Huo, Siyu and Liu, Yuzhou and Chi, Haoen and Yu, Gan and Shi, Xiaoying and Zhu, Wenguang and others},
  journal={Nano Letters},
  volume={25},
  number={45},
  pages={16162--16168},
  year={2025},
  publisher={ACS Publications}
}

@article{likforman1980transitions,
  title={Transitions de la forme de haute temp{\'e}rature $\alpha$ de In2Se3, de part et d'autre de la temp{\'e}rature ambiante},
  author={Likforman, Anna and Fourcroy, Paul-Henri and Guittard, Micheline and Flahaut, Jean and Poirier, Raymond and Szydlo, Nicolas},
  journal={Journal of Solid State Chemistry},
  volume={33},
  number={1},
  pages={91--97},
  year={1980},
  publisher={Elsevier}
}

@article{osamura1966crystal,
  title={Crystal Structures of $\alpha$-and $\beta$-indium selenide, In2Se3},
  author={Osamura, Kozo and Murakami, Yotaro and Tomiie, Yujiro},
  journal={Journal of the Physical Society of Japan},
  volume={21},
  number={9},
  pages={1848--1848},
  year={1966},
  publisher={The Physical Society of Japan}
}

@article{yan2022giant,
  title={Giant tunneling magnetoresistance and electroresistance in $\alpha$-In 2 Se 3-based van der Waals multiferroic tunnel junctions},
  author={Yan, Zhi and Li, Zeyu and Han, Yulei and Qiao, Zhenhua and Xu, Xiaohong},
  journal={Physical Review B},
  volume={105},
  number={7},
  pages={075423},
  year={2022},
  publisher={APS}
}

@article{xue2020optoelectronic,
  title={Optoelectronic ferroelectric domain-wall memories made from a single van der waals ferroelectric},
  author={Xue, Fei and He, Xin and Liu, Wenhao and Periyanagounder, Dharmaraj and Zhang, Chenhui and Chen, Mingguang and Lin, Chun-Ho and Luo, Linqu and Yengel, Emre and Tung, Vincent and others},
  journal={Advanced Functional Materials},
  volume={30},
  number={52},
  pages={2004206},
  year={2020},
  publisher={Wiley Online Library}
}

@article{jacobs2014extraordinary,
  title={Extraordinary photoresponse in two-dimensional In2Se3 nanosheets},
  author={Jacobs-Gedrim, Robin B and Shanmugam, Mariyappan and Jain, Nikhil and Durcan, Christopher A and Murphy, Michael T and Murray, Thomas M and Matyi, Richard J and Moore, Richard L and Yu, Bin},
  journal={ACS nano},
  volume={8},
  number={1},
  pages={514--521},
  year={2014},
  publisher={ACS Publications}
}

@article{popovic1979revised,
  title={Revised and new crystal data for indium selenides},
  author={Popovi{\'c}, Stanko and Tonejc, Antun and Gr{\v{z}}eta-Plenkovi{\'c}, Biserka and {\v{C}}elustka, Branko and Trojko, Rudolf},
  journal={Applied Crystallography},
  volume={12},
  number={4},
  pages={416--420},
  year={1979},
  publisher={International Union of Crystallography}
}

@article{zhou2017out,
  title={Out-of-plane piezoelectricity and ferroelectricity in layered $\alpha$-In2Se3 nanoflakes},
  author={Zhou, Yu and Wu, Di and Zhu, Yihan and Cho, Yujin and He, Qing and Yang, Xiao and Herrera, Kevin and Chu, Zhaodong and Han, Yu and Downer, Michael C and others},
  journal={Nano letters},
  volume={17},
  number={9},
  pages={5508--5513},
  year={2017},
  publisher={ACS Publications}
}

@article{huang2022two,
  title={Two-dimensional In2Se3: A rising advanced material for ferroelectric data storage},
  author={Huang, Yu-Ting and Chen, Nian-Ke and Li, Zhen-Ze and Wang, Xue-Peng and Sun, Hong-Bo and Zhang, Shengbai and Li, Xian-Bin},
  journal={InfoMat},
  volume={4},
  number={8},
  pages={e12341},
  year={2022},
  publisher={Wiley Online Library}
}

@article{kou2020catalyst,
  title={Catalyst-free growth of dense $\gamma$-In2Se3 nanosheet arrays and their application in photoelectric detectors},
  author={Kou, Yumeng and Chen, Lida and Mu, Jianglong and Miao, Hui and Wang, Yishan and Hu, Xiaoyun and Teng, Feng},
  journal={Nanotechnology},
  volume={31},
  number={19},
  pages={195601},
  year={2020},
  publisher={IOP Publishing}
}

@article{shi2020ultrafast,
  title={Ultrafast electrochemical synthesis of defect-free In2Se3 flakes for large-area optoelectronics},
  author={Shi, Huanhuan and Li, Mengmeng and Shaygan Nia, Ali and Wang, Mingchao and Park, SangWook and Zhang, Zhen and Lohe, Martin R and Yang, Sheng and Feng, Xinliang},
  journal={Advanced Materials},
  volume={32},
  number={8},
  pages={1907244},
  year={2020},
  publisher={Wiley Online Library}
}

@article{cui2017significantly,
  title={Significantly enhanced thermoelectric performance of $\gamma$-In2Se3 through lithiation via chemical diffusion},
  author={Cui, Jiaolin and Peng, Hua and Song, Zhiliang and Du, Zhengliang and Chao, Yimin and Chen, Gang},
  journal={Chemistry of Materials},
  volume={29},
  number={17},
  pages={7467--7474},
  year={2017},
  publisher={ACS Publications}
}

@article{cui2015high,
  title={High thermoelectric performance of a defect in $\alpha$-In 2 Se 3-based solid solution upon substitution of Zn for In},
  author={Cui, Jiaolin and Wang, Li and Du, Zhengliang and Ying, Pengzhan and Deng, Yuan},
  journal={Journal of Materials Chemistry C},
  volume={3},
  number={35},
  pages={9069--9075},
  year={2015},
  publisher={Royal Society of Chemistry}
}

@article{wang2023towards,
  title={Towards two-dimensional van der Waals ferroelectrics},
  author={Wang, Chuanshou and You, Lu and Cobden, David and Wang, Junling},
  journal={Nature Materials},
  volume={22},
  number={5},
  pages={542--552},
  year={2023},
  publisher={Nature Publishing Group UK London}
}

@article{liu20192d,
  title={2D materials for quantum information science},
  author={Liu, Xiaolong and Hersam, Mark C},
  journal={Nature Reviews Materials},
  volume={4},
  number={10},
  pages={669--684},
  year={2019},
  publisher={Nature Publishing Group UK London}
}

@article{xia2014two,
  title={Two-dimensional material nanophotonics},
  author={Xia, Fengnian and Wang, Han and Xiao, Di and Dubey, Madan and Ramasubramaniam, Ashwin},
  journal={Nature photonics},
  volume={8},
  number={12},
  pages={899--907},
  year={2014},
  publisher={Nature Publishing Group UK London}
}

@article{liu2020two,
  title={Two-dimensional materials for next-generation computing technologies},
  author={Liu, Chunsen and Chen, Huawei and Wang, Shuiyuan and Liu, Qi and Jiang, Yu-Gang and Zhang, David Wei and Liu, Ming and Zhou, Peng},
  journal={Nature nanotechnology},
  volume={15},
  number={7},
  pages={545--557},
  year={2020},
  publisher={Nature Publishing Group UK London}
}

@article{kupers2018controlled,
  title={Controlled crystal growth of indium selenide, In2Se3, and the crystal structures of $\alpha$-In2Se3},
  author={K\"upers, Michael and Konze, Philipp M and Meledin, Alexander and Mayer, Joachim and Englert, Ulli and Wuttig, Matthias and Dronskowski, Richard},
  journal={Inorganic chemistry},
  volume={57},
  number={18},
  pages={11775--11781},
  year={2018},
  publisher={ACS Publications}
}

@article{kresse1996efficient,
  title={Efficient iterative schemes for ab initio total-energy calculations using a plane-wave basis set},
  author={Kresse, Georg and Furthm{\"u}ller, J{\"u}rgen},
  journal={Physical review B},
  volume={54},
  number={16},
  pages={11169},
  year={1996},
  publisher={APS}
}

@article{kresse1993ab,
  title={Ab initio molecular dynamics for open-shell transition metals},
  author={Kresse, Georg and Hafner, JJPRB},
  journal={Physical Review B},
  volume={48},
  number={17},
  pages={13115},
  year={1993},
  publisher={APS}
  }

@article{kresse1999ultrasoft,
  title={From ultrasoft pseudopotentials to the projector augmented-wave method},
  author={Kresse, Georg and Joubert, Daniel},
  journal={Physical review b},
  volume={59},
  number={3},
  pages={1758},
  year={1999},
  publisher={APS}
}

@article{furness2020accurate,
  title={Accurate and numerically efficient r2SCAN meta-generalized gradient approximation},
  author={Furness, James W and Kaplan, Aaron D and Ning, Jinliang and Perdew, John P and Sun, Jianwei},
  journal={The journal of physical chemistry letters},
  volume={11},
  number={19},
  pages={8208--8215},
  year={2020},
  publisher={ACS Publications}
}

@article{huang2021more,
  title={More Se vacancies in Sb2Se3 under Se-rich conditions: an abnormal behavior induced by defect-correlation in compensated compound semiconductors},
  author={Huang, Menglin and Cai, Zenghua and Wang, Shanshan and Gong, Xin-Gao and Wei, Su-Huai and Chen, Shiyou},
  journal={Small},
  volume={17},
  number={36},
  pages={2102429},
  year={2021},
  publisher={Wiley Online Library}
}

@article{li2020band,
  title={Band structure, ferroelectric instability, and spin--orbital coupling effect of bilayer $\alpha$-In2Se3},
  author={Li, CF and Li, YQ and Tang, YS and Zheng, SH and Zhang, JH and Zhang, Y and Lin, L and Yan, ZB and Jiang, XP and Liu, J-M},
  journal={Journal of Applied Physics},
  volume={128},
  number={23},
  year={2020},
  publisher={AIP Publishing}
}

@article{ding2017prediction,
  title={Prediction of intrinsic two-dimensional ferroelectrics in In2Se3 and other III2-VI3 van der Waals materials},
  author={Ding, Wenjun and Zhu, Jianbao and Wang, Zhe and Gao, Yanfei and Xiao, Di and Gu, Yi and Zhang, Zhenyu and Zhu, Wenguang},
  journal={Nature communications},
  volume={8},
  number={1},
  pages={14956},
  year={2017},
  publisher={Nature Publishing Group UK London}
}

@article{xue2013first,
  title={First-principles study of native point defects in Bi2Se3},
  author={Xue, L and Zhou, P and Zhang, CX and He, CY and Hao, GL and Sun, LZ and Zhong, JX},
  journal={Aip Advances},
  volume={3},
  number={5},
  year={2013},
  publisher={AIP Publishing}
}

@misc{bazantes2026rationalizing,
  title={Rationalizing defect formation energies in metals and semiconductors with semilocal density functionals},
  author={Vega Bazantes, Jorge D and Lebeda, Timo and Ramasamy, Akilan and Pokharel, Kanun and Zhang, Ruiqi and Perdew, John P and Sun, Jianwei},
  year={2026},
  eprint={2604.05385},
  archivePrefix={arXiv},
  url={https://arxiv.org/abs/2604.05385}
}
\end{document}


\title{Supplemental Material for ``Defective $\alpha$-In$_2$Se$_3$ from first principles"}
\author{Jorge Vega Bazantes}
\affiliation{%
Department of Physics and Engineering Physics, Tulane University, New Orleans, Louisiana 70118, USA
}

\author{Timo Lebeda}
\affiliation{Yusuf Hamied Department of Chemistry, University of Cambridge, Cambridge CB2 1EW, UK}

\author{Zhiqiang Mao}
\affiliation{%
Department of Physics, The Pennsylvania State University, University Park, Pennsylvania 16802, USA
}

\author{Jianwei Sun}
\thanks{Contact author: jsun@tulane.edu}
\affiliation{%
Department of Physics and Engineering Physics, Tulane University, New Orleans, Louisiana 70118, USA
}

\maketitle

\section{K points for band structure calculations}

For the band structure calculations we use the the $k$-point path generated by SeeK-path \cite{hinuma2017band,togo2024spglib}. We show below the reciprocal lattice vectors and the high symmetry points used in the calculations.

\begin{table}[h]
\centering
\caption{Reciprocal lattice vectors for the primitive rhombohedral cell of $\alpha$-In$_2$Se$_3$ in units of \AA$^{-1}$.}
\begin{tabular}{c c c c}
\hline\hline
Vector & $x$ & $y$ & $z$ \\
\hline
$\mathbf{b}_1$ & 1.5538 & 0.8971 & 0.2166 \\
$\mathbf{b}_2$ & -1.5538 & 0.8971 & 0.2166 \\
$\mathbf{b}_3$ & 0.0000 & -1.7942 & 0.2166 \\
\hline\hline
\end{tabular}
\label{tab:reciprocal_vectors}
\end{table}

\begin{table}[h]
\centering
\caption{High-symmetry points of the rhombohedral Brillouin zone in fractional reciprocal coordinates.}
\begin{tabular}{c c c c}
\hline\hline
Point & $k_1$ & $k_2$ & $k_3$ \\
\hline
$\Gamma$ & 0.0000 & 0.0000 & 0.0000 \\
$T$      & 0.5000 & 0.5000 & 0.5000 \\
$H_2$    & 0.8236 & 0.1764 & 0.5000 \\
$H_0$    & 0.5000 & -0.1764 & 0.1764 \\
$L$      & 0.5000 & 0.0000 & 0.0000 \\
$S_0$    & 0.3382 & -0.3382 & 0.0000 \\
$S_2$    & 0.6618 & 0.0000 & 0.3382 \\
$F$      & 0.5000 & 0.0000 & 0.5000 \\
\hline\hline
\end{tabular}
\label{tab:high_symmetry_points}
\end{table}

\section{Equations to compute defect formation energies}
To compute the formation energies of the defects, we follow Refs. \cite{hashibon2011first, xue2013first} and \cite{freysoldt2014first}. First we derive the equations as a function of $\Delta \mu_{Se}$, and then do the same for the equations as a function of  $\Delta \mu_{In}$. In the main text we used $E_f$ to denote defect formation energy, here we use $H_f$.

For a defect $X$ at charge $q$, the defect formation energy $H_f[X,q]$ is given by:
\begin{equation}
    H_f[X,q] = E^\mathrm{defective}[X,q]- E^\mathrm{pristine} - \sum_{i} \Delta n_i \mu_i + q E_F + E_{corr} \label{eq:Hf_definition}
\end{equation}
where $E^\mathrm{defective}[X,q]$ is the total energy of the defective supercell, $E^\mathrm{pristine}$ is the total energy of the pristine supercell, $\Delta n_i$ corresponds to the number of atoms of species $i$ either taken or added (for instance, if a single vacancy is formed, then $\Delta n_i = -1$, or if an atom is added then $\Delta n_i = 1$), $\mu_i$ is the chemical potential, which corresponds to the total energy of the elemental phase of species $i$. The chemical potential represents energy of the reservoir from where an atom is exchanged \cite{freysoldt2014first}. $E_F$ is the Fermi energy and represents the energy of the electron reservoir with which the defect exchanges $q$ electrons.
Furthermore, the chemical potentials of In and Se satisfy
\begin{equation}
    \mu_{In_{2}Se_{3}}^\mathrm{bulk} = 2 \mu_{In} + 3 \mu_{Se}\, ,
    \label{eq:mu_bulk_In2Se3}
\end{equation} 

where $\mu_{In_{2}Se_{3}}^\mathrm{bulk}$ is the chemical potential of one formula unit of In$_2$Se$_3$. On the other hand, the chemical potentials of the elemental bulk phases of Se and In are $\mu_{Se}^{0}$ and $\mu_{In}^{0}$ respectively. In our work we take as references the hexagonal Se (P3$_1$21$\bar{1}$, 152) and the tetragonal In (I4/mmm, 139).

We can also define the differences of chemical potential:
\begin{align}
    \Delta \mu_{In} &= \mu_{In} - \mu_{In}^{0}
    \label{eq:delta_mu_in}\\
    \Delta \mu_{Se} &= \mu_{Se} - \mu_{Se}^{0}
    \label{eq:delta_mu_se}\,.
\end{align}

Then the Gibbs free energy to form one formula unit of In$_2$Se$_3$ is given by:
\begin{equation}
    \Delta G (\text{In$_2$Se$_3$}) = \mu_{In_2Se_3}^\mathrm{bulk} - 2 \mu_{In}^{0} - 3\mu_{Se}^{0}\label{eq:Gibbs_free_E} \,.
\end{equation}

Substitute equation \ref{eq:mu_bulk_In2Se3} into equation \ref{eq:Gibbs_free_E} and we get:
\begin{align}
    \Delta G &= 2 \mu_{In} + 3 \mu_{Se} - 2 \mu_{In}^{0} - 3\mu_{Se}^{0}\\
    \Delta G &= 2(\mu_{In} -  \mu_{In}^{0}) + 3 (\mu_{Se} - \mu_{Se}^{0})\\
\end{align}

Then replacing \ref{eq:delta_mu_in} and \ref{eq:delta_mu_se} in the previous equation we get:

\begin{align}
    \Delta G &= 2\Delta \mu_{In} + 3 \Delta\mu_{Se} \label{eq:DeltaG_deltas}
\end{align}

Next, we derive two equation as a function of $\Delta \mu_{Se}$ and $\Delta \mu_{In}$ using equations \ref{eq:Hf_definition}, \ref{eq:mu_bulk_In2Se3} and \ref{eq:Gibbs_free_E}. \\
First let us consider equation \ref{eq:Hf_definition} for neutral defects, i.e., $q=0$ and $E_{corr}=0$:
\begin{align}
    H_f[X,q] &= E^\mathrm{defective}[X,q]- E^\mathrm{pristine} - \sum_{i} \Delta n_i \mu_i \label{eq:Hf_neutral}\\
    H_f[X,q] &= E^\mathrm{defective}[X,q]- E^\mathrm{pristine} - (\Delta n_{In}\mu_{In} +\Delta n_{Se}\mu_{Se})\label{eq:Hf_expanded}
\end{align}

We can rewrite equation \ref{eq:mu_bulk_In2Se3} to replace into $\mu_{In}$ and $\mu_{Se}$. For clarity let us consider only the last two terms of the previous equation:
\begin{align}
    \Delta n_{In}\mu_{In} +\Delta n_{Se}\mu_{Se} &= \Delta n_{In}\ \bigg(\frac{\mu_{In_2Se_3}^\mathrm{bulk} - 3\mu_{Se}}{2} \bigg) + 
    \Delta n_{Se} \mu_{Se}\label{eq:deltas_dec_1}
\end{align}
Using equation \ref{eq:Gibbs_free_E} for the Gibbs free energy, we can re-write:
\begin{equation}
        \mu_{In_2Se_3}^\mathrm{bulk}  = \Delta G (\text{In$_2$Se$_3$}) + 2 \mu_{In}^{0} + 3\mu_{Se}^{0}\label{eq:mu_bulk(G)}
\end{equation}

and replacing into equation \ref{eq:deltas_dec_1}, we get:

\begin{align}
   \Delta n_{In}\mu_{In} +\Delta n_{Se}\mu_{Se} &= 
   \frac{\Delta n_{In}}{2}( \Delta G + 2 \mu_{In}^{0} + 3\mu_{Se}^{0} - 3\mu_{Se}) + 
    \Delta n_{Se}\mu_{Se}\\
    &=   \frac{\Delta n_{In}}{2}( \Delta G + 2 \mu_{In}^{0} + 3(\mu_{Se}^{0} -\mu_{Se})) + 
    \Delta n_{Se}\mu_{Se}\\
    &= \frac{\Delta n_{In}}{2}( \Delta G + 2 \mu_{In}^{0} - 3(\Delta \mu_{Se})) + 
    \Delta n_{Se}\mu_{Se}\\
    &= \frac{\Delta n_{In}}{2}( \Delta G + 2 \mu_{In}^{0} - 3(\Delta \mu_{Se})) + 
    \Delta n_{Se}\mu_{Se}\\
    &= \bigg(\frac{\Delta n_{In} \Delta G}{2} + \Delta n_{In} \mu_{In}^{0} - \frac{3}{2}\Delta n_{In}\Delta \mu_{Se} \bigg) + \Delta n_{Se}\mu_{Se}\label{eq:delta_dec_2}
\end{align}

Then, we replace equation \ref{eq:delta_dec_2} into \ref{eq:Hf_expanded}, $H_f[X,q]$:
\begin{align}
    H_f[X,q] &= E^\mathrm{defective}[X,q]- E^\mathrm{pristine} - (\Delta n_{In}\mu_{In} +\Delta n_{Se}\mu_{Se})\\
    &=E^\mathrm{defective}[X,q]- E^\mathrm{pristine} - \frac{\Delta n_{In} \Delta G}{2} - \Delta n_{In} \mu_{In}^{0} + \frac{3}{2}\Delta n_{In}\Delta \mu_{Se} - \Delta n_{Se}\mu_{Se}
\end{align}

where we can define:
\begin{align}
    \Delta n_{Se}\mu_{Se} &= \Delta n_{Se} (\mu_{Se} + \mu_{Se}^{0} - \mu_{Se}^{0})\\
    &=  \Delta n_{Se} ((\mu_{Se} - \mu_{Se}^{0}) + \mu_{Se}^{0})\\
    &=  \Delta n_{Se} (\Delta \mu_{Se} + \mu_{Se}^{0})\\
    &= \Delta n_{Se}\Delta \mu_{Se} + \Delta n_{Se} \mu_{Se}^{0}
\end{align}

Hence we have
\begin{align}
    H_f[X,q] &= E^\mathrm{defective}[X,q]- E^\mathrm{pristine} - (\Delta n_{In}\mu_{In} +\Delta n_{Se}\mu_{Se})\\
    &=E^\mathrm{defective}[X,q]- E^\mathrm{pristine} - \frac{\Delta n_{In} \Delta G}{2} - \Delta n_{In} \mu_{In}^{0} + \frac{3}{2}\Delta n_{In}\Delta \mu_{Se} - \Delta n_{Se}\mu_{Se}\\
    &=E^\mathrm{defective}[X,q]- E^\mathrm{pristine} - \frac{\Delta n_{In} \Delta G}{2} - \Delta n_{In} \mu_{In}^{0} + \frac{3}{2}\Delta n_{In}\Delta \mu_{Se} - (\Delta n_{Se}\Delta \mu_{Se} + \Delta n_{Se} \mu_{Se}^{0})
\end{align}

Let us define $H_f^{0}$:
\begin{equation}
H_f^{0} = E^\mathrm{defective}[X,q]- E^\mathrm{pristine} - \Delta n_{In}\mu_{In}^{0} -\Delta n_{Se} \mu_{Se}^{0} \,.
\end{equation}

Finally, we get the defect formation energy as a function of $\Delta \mu_{Se}$:
\begin{align}
    H_f[X,q] &= H_f^{0} - \frac{1}{2}\Delta n_{In} \Delta G - 
    \Delta \mu_{Se}\big(\Delta n_{Se}-  \frac{3}{2}\Delta n_{In}\big)\label{eq:Ef_mu_Se}
\end{align}

\clearpage
\subsection{Se-rich environment: $\Delta \mu Se = 0$}
For Se rich environment we have $\Delta \mu Se = 0$. For In rich environment we have $\Delta \mu In = 0$. 

The defect formation energy as a function of $\Delta \mu_{Se}$ is given by equation \ref{eq:Ef_mu_Se}. If we set  $\Delta \mu_{Se}=0$, we get: 

\begin{align}
    H_f[X] &= H_f^{0} - \frac{1}{2}\Delta n_{In} \Delta G \\
             &= E^\mathrm{defective}- E^\mathrm{pristine} - \Delta n_{In}\mu_{In}^{0} -\Delta n_{Se} \mu_{Se}^{0} -\frac{1}{2}\Delta n_{In} \Delta G
\end{align}

\subsubsection{Se vacancy}
For a  Se vacancy we have no change in the In atoms ($\Delta n_{In} = 0$) and since one Se atom is taken out, we have $\Delta n_{Se} = -1$, hence the formation energy is given by:
\begin{equation}
    H_f[X,q] = E^\mathrm{defective} - E^\mathrm{pristine} + \mu_{Se}^{0} 
\end{equation}

where $E^\mathrm{defective}$ is the total energy of the defective supercell, $E^\mathrm{pristine}$ is the total energy of the pristine non defective supercell, and $\mu_{Se}^{0}$ is the chemical potential of the ideal bulk Se, which is the total energy of a bulk calculation of Se.

\subsubsection{In vacancy}
For an In vacancy, one In atom is taken out ($\Delta n_{In} = -1$) and the number of Se atoms remain the same ($\Delta n_{Se} = 0$). Then, the formation energy is given by:

\begin{align}
    H_{f} &= E_f^0 - \frac{1}{2}\Delta G \Delta n_{In}  
    \\
    &= E^\mathrm{defective} - E^\mathrm{pristine} + \mu_{In}^{0} + \frac{1}{2}\Delta G 
\end{align}

where the Gibbs free energy $\Delta G = \mu_{In_2Se_3}^\mathrm{bulk} - 2\mu_{In}^{0} - 3\mu_{Se}^0$. Hence,
\begin{align}
    H_{f} &= E^\mathrm{defective} - E^\mathrm{pristine} + \mu_{In}^{0} + \frac{1}{2}(\mu_{In_2Se_3}^\mathrm{bulk} - 2\mu_{In}^{0} - 3\mu_{Se}^0) \\
    &= E^\mathrm{defective} - E^\mathrm{pristine} + \mu_{In}^{0} + \frac{1}{2}\mu_{In_2Se_3}^\mathrm{bulk} - \mu_{In}^{0} - \frac{3}{2}\mu_{Se}^0 \\
    &= E^\mathrm{defective} - E^\mathrm{pristine} + \frac{1}{2}\mu_{In_2Se_3}^\mathrm{bulk} - \frac{3}{2}\mu_{Se}^0 
\end{align}

\subsubsection{Antisite $Se_{In}$, Se replaces In}
For the $Se_{In}$ antisite, $\Delta n_{In} = -1$ and $\Delta n_{Se} = +1$. Hence,
\begin{align}
    H_{f} &= E_f^0 - \frac{1}{2}\Delta G \Delta n_{In}  
    \\
    &= E^\mathrm{defective}- E^\mathrm{pristine} - \Delta n_{In}\mu_{In}^{0} -\Delta n_{Se} \mu_{Se}^{0} -\frac{1}{2}\Delta n_{In} \Delta G\\
    &=E^\mathrm{defective}- E^\mathrm{pristine} + \mu_{In}^{0} - \mu_{Se}^{0} +\frac{1}{2} \Delta G\\
    &= E^\mathrm{defective}- E^\mathrm{pristine} + \mu_{In}^{0} - \mu_{Se}^{0} +\frac{1}{2} (\mu_{In_2Se_3}^\mathrm{bulk} - 2\mu_{In}^{0} - 3\mu_{Se}^0)\\
    &= E^\mathrm{defective}- E^\mathrm{pristine} -\frac{5}{2}\mu_{Se}^{0} + \frac{1}{2}\mu_{In_2Se_3}^\mathrm{bulk}
    \end{align}





\subsubsection{Antisite $In_{Se}$, In replaces Se}
For the $In_{Se}$ antisite, $\Delta n_{In} = +1$ and $\Delta n_{Se} = -1$. Hence, we have:
\begin{align}
    H_{f} &= H_f^0 - \frac{1}{2}\Delta G \Delta n_{In}  
    \\
    &= E^\mathrm{defective} - E^\mathrm{pristine} - \Delta n_{In} \mu_{In}^{0} - \Delta n_{Se} \mu_{Se}^{0} - \frac{1}{2}\Delta G \Delta n_{In} \\
    &= E^\mathrm{defective} - E^\mathrm{pristine} -  \mu_{In}^{0} + \mu_{Se}^{0} - \frac{1}{2}\Delta G \\
    &= E^\mathrm{defective} - E^\mathrm{pristine} -  \mu_{In}^{0} + \mu_{Se}^{0} - \frac{1}{2}(\mu_{In_2Se_3}^\mathrm{bulk} - 2\mu_{In}^{0} - 3\mu_{Se}^0)\\
    &= E^\mathrm{defective} - E^\mathrm{pristine} + \frac{5}{2}\mu_{Se}^{0} - \frac{1}{2}\mu_{In_2Se_3}^\mathrm{bulk}
\end{align}

\subsection{In-rich condition: $\Delta \mu In = 0$}
To get the equations for the In-rich condition, we need to derive the equation of defect formation energy as a function of $\Delta \mu_{In}$. Hence as done before, we derive the final expression using equations \ref{eq:Hf_definition}, \ref{eq:mu_bulk_In2Se3} and \ref{eq:Gibbs_free_E}. \\
First let us consider equation \ref{eq:Hf_definition} for neutral defects, i.e., $q=0$ and $E_{corr}=0$:
\begin{align}
    H_f[X,q] &= E^\mathrm{defective}[X,q]- E^\mathrm{pristine} - \sum_{i} \Delta n_i \mu_i \label{eq:Hf_neutral_In}\\
    H_f[X,q] &= E^\mathrm{defective}[X,q]- E^\mathrm{pristine} - (\Delta n_{In}\mu_{In} +\Delta n_{Se}\mu_{Se})\label{eq:Hf_expanded_In}
\end{align}

We can rewrite equation \ref{eq:mu_bulk_In2Se3} to replace into $\mu_{In}$ and $\mu_{Se}$. For clarity let us consider only the last two terms of the previous equation:
\begin{align}
    \Delta n_{In}\mu_{In} +\Delta n_{Se}\mu_{Se} &=  \Delta n_{In}\mu_{In}  + 
    \Delta n_{Se} \bigg(
    \frac{\mu_{In_2Se_3}^\mathrm{bulk}-2 \mu_{In}}{3}\bigg)\label{eq:deltas_dec_1_In}
\end{align}
Using equation \ref{eq:Gibbs_free_E} for the Gibbs free energy, we can re-write:
\begin{equation}
        \mu_{In_2Se_3}^\mathrm{bulk}  = \Delta G (\text{In$_2$Se$_3$}) + 2 \mu_{In}^{0} + 3\mu_{Se}^{0}\label{eq:mu_bulk(G)_In}
\end{equation}

and replacing into equation \ref{eq:deltas_dec_1_In}, we get:

\begin{align}
   \Delta n_{In}\mu_{In} +\Delta n_{Se}\mu_{Se} &=
   \Delta n_{In}\mu_{In} +
   \Delta n_{Se} \bigg(\frac{\mu_{In_2Se_3}^\mathrm{bulk}-2 \mu_{In}}{3}\bigg)\\
   &=  \Delta n_{In}\mu_{In} +  \Delta n_{Se} \bigg(\frac{\Delta G + 2\mu_{In}^{0} + 3\mu_{Se}^{0} - 2\mu_{In}}{3}\bigg)\\
   &= \Delta n_{In}\mu_{In} +  \Delta n_{Se} \bigg(\frac{\Delta G + 2(\mu_{In}^{0} - \mu_{In}) + 3\mu_{Se}^{0} }{3}\bigg)\\
   &= \Delta n_{In}\mu_{In} +  \Delta n_{Se} \bigg(\frac{\Delta G - 2(\Delta\mu_{In}) + 3\mu_{Se}^{0} }{3}\bigg)\\
   &= \Delta n_{In}\mu_{In} +  \frac{1}{3}\Delta n_{Se}\Delta G - \frac{2}{3}\Delta n_{Se} \Delta \mu_{In} + \Delta n_{Se}\mu_{Se}^{0} 
  \label{eq:delta_dec_2_In}
\end{align}

Then, we replace equation \ref{eq:delta_dec_2_In} into \ref{eq:Hf_expanded_In}, $H_f[X,q]$:
\begin{align}
    H_f[X,q] &= E^\mathrm{defective}- E^\mathrm{pristine} - (\Delta n_{In}\mu_{In} +\Delta n_{Se}\mu_{Se})\\
    &=E^\mathrm{defective}- E^\mathrm{pristine} - (\Delta n_{In}\mu_{In} +  \frac{1}{3}\Delta n_{Se}\Delta G - \frac{2}{3}\Delta n_{Se} \Delta \mu_{In} + \Delta n_{Se}\mu_{Se}^{0})
\end{align}

where we can define:
\begin{align}
    \Delta n_{In}\mu_{In} &= \Delta n_{In} (\mu_{In} + \mu_{In}^{0} - \mu_{In}^{0})\\
    &=  \Delta n_{In} ((\mu_{In} - \mu_{In}^{0}) + \mu_{In}^{0})\\
    &=  \Delta n_{In} (\Delta \mu_{In} + \mu_{In}^{0})\\
    &= \Delta n_{In}\Delta \mu_{In} + \Delta n_{In} \mu_{In}^{0}
\end{align}

Hence we have
\begin{align}
    H_f[X,q] &= E^\mathrm{defective}- E^\mathrm{pristine} - (\Delta n_{In}\mu_{In} +\Delta n_{Se}\mu_{Se})\\
    &=E^\mathrm{defective}- E^\mathrm{pristine} - (
    \Delta n_{In}\Delta \mu_{In} 
    + \Delta n_{In} \mu_{In}^{0} 
    +  \frac{1}{3}\Delta n_{Se}\Delta G 
    - \frac{2}{3}\Delta n_{Se} \Delta \mu_{In} 
    + \Delta n_{Se}\mu_{Se}^{0})\\
    &= E^\mathrm{defective}- E^\mathrm{pristine} 
    - \Delta n_{In}\Delta \mu_{In}
    - \Delta n_{In} \mu_{In}^{0} 
    - \frac{1}{3}\Delta n_{Se}\Delta G 
    + \frac{2}{3}\Delta n_{Se} \Delta \mu_{In} 
    - \Delta n_{Se}\mu_{Se}^{0}\\
    &= (E^\mathrm{defective}- E^\mathrm{pristine} 
    - \Delta n_{In} \mu_{In}^{0} 
    - \Delta n_{Se}\mu_{Se}^{0})
    - \Delta n_{In}\Delta \mu_{In}
    - \frac{1}{3}\Delta n_{Se}\Delta G 
    + \frac{2}{3}\Delta n_{Se} \Delta \mu_{In}\\
    &= (E^\mathrm{defective}- E^\mathrm{pristine} 
    - \Delta n_{In} \mu_{In}^{0} 
    - \Delta n_{Se}\mu_{Se}^{0})
    - \Delta \mu_{In} (\Delta n_{In} - \frac{2}{3}\Delta n_{Se} )
    - \frac{1}{3}\Delta n_{Se}\Delta G\\
    &= H_{f}^{0}
    - \frac{1}{3}\Delta n_{Se}\Delta G
    - \Delta \mu_{In} (\Delta n_{In} - \frac{2}{3}\Delta n_{Se} )
\end{align}

where we have defined:
\begin{equation}
H_f^{0} = E^\mathrm{defective}- E^\mathrm{pristine} - \Delta n_{In}\mu_{In}^{0} -\Delta n_{Se} \mu_{Se}^{0} \,.
\end{equation}

Finally, we get the defect formation energy as a function of $\Delta \mu_{In}$:
\begin{align}
H_f &= H_{f}^{0}
    - \frac{1}{3}\Delta n_{Se}\Delta G
    - \Delta \mu_{In} \bigg(\Delta n_{In} - \frac{2}{3}\Delta n_{Se} \bigg)
\end{align}

Then, as done before with the Se-rich conditions, we derive the equations for four types of defects, however now for In-rich conditions we have $\Delta\mu_{In}=0$, hence we have:
\begin{align}
H_f &= H_{f}^{0}
    - \frac{1}{3}\Delta n_{Se}\Delta G
    - \Delta \mu_{In} \bigg(\Delta n_{In} - \frac{2}{3}\Delta n_{Se} \bigg)\\
    &= H_{f}^{0}
    - \frac{1}{3}\Delta n_{Se}\Delta G\\
    &= E^\mathrm{defective}- E^\mathrm{pristine} - \Delta n_{In}\mu_{In}^{0} -\Delta n_{Se} \mu_{Se}^{0} 
    - \frac{1}{3}\Delta n_{Se}\Delta G \label{eq:Hf_In_rich}   
\end{align}

Hence we use equation \ref{eq:Hf_In_rich} for obtaining the final equations.
\subsubsection{Se vacancy}
For a  Se vacancy we have no change in the In atoms ($\Delta n_{In} = 0$) and since one Se atom is taken out, we have $\Delta n_{Se} = -1$, hence the formation energy is given by:
\begin{align}
H_f &= E^\mathrm{defective}- E^\mathrm{pristine} - \Delta n_{In}\mu_{In}^{0} -\Delta n_{Se} \mu_{Se}^{0} 
    - \frac{1}{3}\Delta n_{Se}\Delta G\\
    &= E^\mathrm{defective}- E^\mathrm{pristine} + \mu_{Se}^{0} + \frac{1}{3}\Delta G\\
    &= E^\mathrm{defective}- E^\mathrm{pristine} + \mu_{Se}^{0} + \frac{1}{3}(\mu_{In_2Se_3}^\mathrm{bulk} - 2\mu_{In}^{0} - 3\mu_{Se}^{0})\\
    &= E^\mathrm{defective}- E^\mathrm{pristine} + \frac{1}{3}\mu_{In_2Se_3}^\mathrm{bulk} - \frac{2}{3}\mu_{In}^{0}
\end{align}

\subsubsection{In vacancy}
For an In vacancy, one In atom is taken out ($\Delta n_{In} = -1$) and the number of Se atoms remain the same ($\Delta n_{Se} = 0$). Then, the formation energy is given by:
\begin{align}
H_f &= E^\mathrm{defective}- E^\mathrm{pristine} - \Delta n_{In}\mu_{In}^{0} -\Delta n_{Se} \mu_{Se}^{0} 
    - \frac{1}{3}\Delta n_{Se}\Delta G\\
    &= E^\mathrm{defective}- E^\mathrm{pristine} + \mu_{In}^{0}
\end{align}

\subsubsection{Antisite $Se_{In}$, Se replaces In}
For the $Se_{In}$ antisite, $\Delta n_{In} = -1$ and $\Delta n_{Se} = +1$. Hence,
\begin{align}
H_f &= E^\mathrm{defective}- E^\mathrm{pristine} - \Delta n_{In}\mu_{In}^{0} -\Delta n_{Se} \mu_{Se}^{0} 
    - \frac{1}{3}\Delta n_{Se}\Delta G\\
    &= E^\mathrm{defective}- E^\mathrm{pristine} + \mu_{In}^{0} - \mu_{Se}^{0}-\frac{1}{3}(\mu_{In_2Se_3}^\mathrm{bulk} - 2\mu_{In}^{0} - 3\mu_{Se}^{0})\\
    &= E^\mathrm{defective}- E^\mathrm{pristine} + \mu_{In}^{0}  - \frac{1}{3}\mu_{In_2Se_3}^\mathrm{bulk} + \frac{2}{3}\mu_{In}^{0}\\
    &= E^\mathrm{defective}- E^\mathrm{pristine} +
    \frac{5}{3}\mu_{In}^{0}-\frac{1}{3}\mu_{In_2Se_3}^\mathrm{bulk}
\end{align}

\subsubsection{Antisite $In_{Se}$, In replaces Se}
For the $In_{Se}$ antisite, $\Delta n_{In} = +1$ and $\Delta n_{Se} = -1$. Hence, we have:
\begin{align}
H_f &= E^\mathrm{defective}- E^\mathrm{pristine} - \Delta n_{In}\mu_{In}^{0} -\Delta n_{Se} \mu_{Se}^{0} 
    - \frac{1}{3}\Delta n_{Se}\Delta G\\
    &= E^\mathrm{defective}- E^\mathrm{pristine} -\mu_{In}^{0}+\mu_{Se}^{0}+\frac{1}{3}(\mu_{In_2Se_3}^\mathrm{bulk} - 2\mu_{In}^{0} - 3\mu_{Se}^{0})\\
    &= E^\mathrm{defective}- E^\mathrm{pristine} - \frac{5}{3}\mu_{In}^{0} + \frac{1}{3}\mu_{In_2Se_3}^\mathrm{bulk}
\end{align}

\clearpage





\section{Defect formation energies}
In this section we present the numbers obtained for defect formation energies. 

\begin{table}[h!]
\centering
\caption{Defect formation energies of vacancies in Se and In-rich environments. The energies are also depicted in Figure 5 in the main text.}
\begin{tabular}{cccc}
\hline\hline
Defect type & Layer positioning & Se-rich (eV) & In-rich (eV) \\
\hline
Se vacancy 1 ($V_{Se1}$)  & Upper   & 2.46 & 1.34 \\
Se vacancy 2 ($V_{Se2}$)  & Middle  & 1.45 & 0.33 \\
Se vacancy 3 ($V_{Se3}$)  & Bottom  & 2.56 & 1.44 \\
In vacancy 1 ($V_{In1}$)  & Upper   & 2.57 & 4.25 \\
In vacancy 2 ($V_{In2}$)  & Lower   & 3.12 & 4.80 \\
\hline\hline                       
\end{tabular}
\label{vacancies}
\end{table}

\begin{table}[h!]
\centering
\caption{Defect formation energies of antisite defects in Se and In-rich environments. The energies are depicted in Figure 5.}
\begin{tabular}{cccc}
\hline\hline
Defect type & Layer positioning & Se-rich (eV) & In-rich (eV) \\
\hline
In replaces Se 1 ($In_{Se1}$)  & Upper & 4.24 & 1.44 \\
In replaces Se 2 ($In_{Se2}$) & Middle & 4.14 & 1.34 \\
In replaces Se 3 ($In_{Se3}$)& Bottom  & 4.66 & 1.86 \\
Se replaces In 1 ($Se_{In1}$) & Upper  & 1.44 & 4.24 \\
Se replaces In 2 ($Se_{In2}$)& Lower   & 1.24 & 4.03 \\
\hline\hline                       
\end{tabular}
\label{antisites}
\label{end}
\end{table}

\begin{table}[h!]
\centering
\caption{Defect formation energies of vacancies computed on the unrelaxed geometry, $E_{f}^{\mathrm{unrelaxed}}$. The chemical potential references, the pristine cell, and the definition of the defect formation energy, Eq.~(1), are the same as in Table~\ref{vacancies}. \mbox{$E_{f}^{\mathrm{relaxation}} = E_{f}^{\mathrm{unrelaxed}} - E_{f}^{\mathrm{relaxed}}$} is the energy gained on relaxation and is independent of the chemical potentials.}
\begin{tabular}{ccccc}
\hline\hline
Defect type & Layer positioning & Se-rich (eV) & In-rich (eV) & $E_{f}^{\mathrm{relaxation}}$ (eV) \\
\hline
Se vacancy 1 ($V_{Se1}$)             & Upper   & 3.04 & 1.93 & 0.59 \\
Se vacancy 2 ($V_{Se2}$)             & Middle  & 2.75 & 1.63 & 1.29 \\
Se vacancy 3 ($V_{Se3}$)             & Bottom  & 2.75 & 1.63 & 0.20 \\
In vacancy 1 ($V_{In1}$)             & Upper   & 2.93 & 4.61 & 0.37 \\
In vacancy 2 ($V_{In2}$)             & Lower   & 3.39 & 5.07 & 0.27 \\
\hline\hline
\end{tabular}
\label{vacancies_unrelaxed}
\end{table}

\begin{table}[h!]
\centering
\caption{Defect formation energies of antisite defects computed on the unrelaxed geometry, $E_{f}^{\mathrm{unrelaxed}}$. The references, the pristine cell, and the definition of the defect formation energy, Eq.~(1), are the same as in Table~\ref{antisites}. \mbox{$E_{f}^{\mathrm{relaxation}} = E_{f}^{\mathrm{unrelaxed}} - E_{f}^{\mathrm{relaxed}}$} is the energy gained on relaxation and is independent of the chemical potentials.}
\begin{tabular}{ccccc}
\hline\hline
Defect type & Layer positioning & Se-rich (eV) & In-rich (eV) & $E_{f}^{\mathrm{relaxation}}$ (eV) \\
\hline
In replaces Se 1 ($In_{Se1}$)        & Upper   & 5.08 & 2.28 & 0.84 \\
In replaces Se 2 ($In_{Se2}$)        & Middle  & 4.85 & 2.05 & 0.71 \\
In replaces Se 3 ($In_{Se3}$)        & Bottom  & 5.26 & 2.46 & 0.60 \\
Se replaces In 1 ($Se_{In1}$)        & Upper   & 2.15 & 4.95 & 0.71 \\
Se replaces In 2 ($Se_{In2}$)        & Lower   & 1.67 & 4.46 & 0.43 \\
\hline\hline
\end{tabular}
\label{antisites_unrelaxed}
\end{table}

\clearpage

\section{Bond lengths and angles of vacancies in $\alpha$-In$_2$Se$_3$}

\begin{table}[hb!]
\centering
\caption{All Se--In--Se bond angles at the In atom that loses a Se neighbor when the vacancy forms: In1 for $V_{Se1}$ and $V_{Se2}$, In2 for $V_{Se3}$. The pristine rows give the same In atom before the vacancy is formed. $N_c$ is the coordination number and $N_{angles}$ the multiplicity of each angle. $V_{Se2}$ is the only center with a single angle value, meaning its remaining Se neighbors are still equivalent to one another.}
\begin{tabular}{cccccc}
\hline\hline
Structure & Center & $N_c$ & Angle & Value (deg) & $N_{angles}$ \\
\hline
pristine  & In1 & 4 & Se1--In1--Se1 & 100.56 & 3 \\
          &     &   & Se1--In1--Se2 & 117.35 & 3 \\
\hline
$V_{Se2}$ & In1 & 3 & Se1--In1--Se1 &  76.06 & 3 \\
\hline
$V_{Se1}$ & In1 & 3 & Se1--In1--Se1 &  92.91 & 1 \\
          &     &   & Se1--In1--Se2 & 110.63 & 2 \\
\hline
pristine  & In2 & 6 & Se2--In2--Se3 &  86.12 & 6 \\
          &     &   & Se2--In2--Se2 &  90.23 & 3 \\
          &     &   & Se3--In2--Se3 &  97.28 & 3 \\
          &     &   & Se2--In2--Se3 & 174.82 & 3 \\
\hline
$V_{Se3}$ & In2 & 5 & Se2--In2--Se3 &  86.26 & 2 \\
          &     &   & Se2--In2--Se2 &  89.09 & 1 \\
          &     &   & Se2--In2--Se3 &  94.13 & 2 \\
          &     &   & Se3--In2--Se3 &  94.36 & 1 \\
          &     &   & Se2--In2--Se2 & 101.23 & 2 \\
          &     &   & Se2--In2--Se3 & 164.55 & 2 \\
\hline\hline
\end{tabular}
\label{all_angles}
\end{table}

\begin{table}[hb!]
\centering
\caption{Distances after forming $V_{Se1}$, in \AA, for the atoms adjacent to the vacancy, with the multiplicity in parentheses. The $V_{Se1}$ row and column give distances to the vacant site, the position the removed Se1 occupied, and are therefore separations, not bonds; they are excluded from $N_c$. In the pristine cell this distance was 2.626~\AA\ to each of the three In1. The In1--In1 entry is likewise excluded from $N_c$: the three In1 left behind contract from 4.040 to 3.382~\AA\ of one another, essentially the second-neighbor separation of 3.385~\AA\ in our relaxed bulk In reference. Because the vacancy makes atoms of the same type inequivalent, each row describes a specific atom, identified in the footnotes, rather than a site type; the matrix is correspondingly not symmetric.}
\begin{tabular}{cccccccc}
\hline\hline
 & Se1 & In1 & $V_{Se1}$ & Se2 & In2 & Se3 & $N_c$ \\
\hline
Se1$^{a}$ & --        & 2.622~(1) & --      & --        & --        & --        & 3 \\
          &           & 2.652~(1) &         &           &           &           &   \\
          &           & 2.682~(1) &         &           &           &           &   \\
In1$^{b}$ & 2.682~(2) & 3.382~(2) & [2.276] & 2.564~(1) & --        & --        & 3 \\
$V_{Se1}$ & --        & [2.276]   & --      & --        & --        & --        & -- \\
Se2$^{c}$ & --        & 2.564~(1) & --      & --        & 2.828~(2) & --        & 4 \\
          &           &           &         &           & 2.901~(1) &           &   \\
In2$^{d}$ & --        & --        & --      & 2.828~(2) & --        & 2.688~(1) & 6 \\
          &           &           &         & 2.870~(1) &           & 2.701~(2) &   \\
Se3$^{e}$ & --        & --        & --      & --        & 2.688~(3) & --        & 3 \\
\hline\hline
\end{tabular}

\begin{flushleft}
\footnotesize
$^{a}$ One of the two Se1 remaining on In1$^{b}$.\\
$^{b}$ One of the three In1 that lost the removed Se1; it displaces 0.382~\AA,
almost entirely in-plane.\\
$^{c}$ The Se2 bonded axially to In1$^{b}$.\\
$^{d}$ An In2 bonded to Se2$^{c}$.\\
$^{e}$ A Se3 on In2$^{d}$.
\end{flushleft}
\label{defect_distances_VSe1}
\end{table}

\begin{table}[h!]
\centering
\caption{Distances after forming $V_{Se3}$, in \AA, for the atoms adjacent to the vacancy, with the multiplicity in parentheses. The $V_{Se3}$ row and column give distances to the vacant site, the position the removed Se3 occupied, and are therefore separations, not bonds; they are excluded from $N_c$. In the pristine cell this distance was 2.691~\AA\ to each of the three In2. Because the vacancy makes atoms of the same type inequivalent, each row describes a specific atom, identified in the footnotes, rather than a site type; the matrix is correspondingly not symmetric.}
\begin{tabular}{cccccccc}
\hline\hline
 & Se1 & In1 & Se2 & In2 & Se3 & $V_{Se3}$ & $N_c$ \\
\hline
Se1$^{a}$ & --        & 2.611~(1) & --        & --        & --        & --      & 3 \\
          &           & 2.626~(2) &           &           &           &         &   \\
In1$^{b}$ & 2.604~(2) & --        & 2.558~(1) & --        & --        & --      & 4 \\
          & 2.611~(1) &           &           &           &           &         &   \\
Se2$^{c}$ & --        & 2.558~(1) & --        & 2.652~(1) & --        & --      & 4 \\
          &           &           &           & 2.907~(2) &           &         &   \\
In2$^{d}$ & --        & --        & 2.652~(1) & --        & 2.679~(2) & [3.115] & 5 \\
          &           &           & 2.794~(2) &           &           &         &   \\
Se3$^{e}$ & --        & --        & --        & 2.679~(1) & --        & --      & 3 \\
          &           &           &           & 2.695~(1) &           &         &   \\
          &           &           &           & 2.715~(1) &           &         &   \\
$V_{Se3}$ & --        & --        & --        & [3.115]   & --        & --      & -- \\
\hline\hline
\end{tabular}

\begin{flushleft}
\footnotesize
$^{a}$ A Se1 bonded to In1$^{b}$.\\
$^{b}$ An In1 bonded to Se2$^{c}$; it is two layers from the vacancy and
essentially unperturbed.\\
$^{c}$ A Se2 bonded to In2$^{d}$.\\
$^{d}$ One of the three In2 that lost the removed Se3; it displaces 0.440~\AA\
away from the vacant site.\\
$^{e}$ One of the two Se3 remaining on In2$^{d}$.
\end{flushleft}
\label{defect_distances_VSe3}
\end{table}


\section{Bond-valence analysis of the Se sites}
\label{sm:bond_valence}

\subsection{Definitions}

The bond-valence model~\cite{brown1985bond, brese1991bond} assigns to every
cation--anion bond a \emph{bond valence} $s$ that decreases exponentially with
the bond length $d$,
\begin{equation}
s_{ij} \;=\; \exp\!\left[\frac{R_{0}^{ij} - d_{ij}}{b}\right],
\label{eq:sm_bv}
\end{equation}
where $i$ labels the cation, $j$ the anion, $R_{0}^{ij}$ is an empirical
parameter specific to the element pair, and $b=0.37$~\AA\ is the universal
softness parameter recommended in Ref.~\cite{brese1991bond}. Valences are quoted in
valence units (v.u.). The \emph{bond valence sum} at a site,
\begin{equation}
V_{i} \;=\; \sum_{j} s_{ij},
\label{eq:sm_bvs}
\end{equation}
runs over all bonds at that site and should approximate the formal oxidation
state; for a Se site in these compounds the expected value is $|{-2}|=2$~v.u.
Throughout we use $R_{0}(\text{In--Se})=2.47$~\AA\ and
$R_{0}(\text{Bi--Se})=2.72$~\AA, both taken from Table~3 of
Ref.~\cite{brese1991bond}, which tabulates anions other than O, F and Cl.
Equation~(\ref{eq:sm_bv}) is evaluated with the cation--anion distances of the
relaxed r$^2$SCAN+rVV10 structures; no fitting of any kind is performed.

\subsection{Worked example: the $Se2$ site of $\alpha$-In$_2$Se$_3$-3R}

The $Se2$ site is the only four-fold coordinated Se site in the structure. In
the relaxed supercell it has one axial $In1$ neighbor at $d=2.5360$~\AA\ and
three basal $In2$ neighbors at $d=2.8508$~\AA. Equation~(\ref{eq:sm_bv}) gives,
for the axial bond,
\begin{equation}
\begin{split}
s(In1\text{--}Se2) &= \exp\!\left[\frac{2.47 - 2.5360}{0.37}\right]\\
                   &= \exp(-0.1784) = 0.8366~\text{v.u.},
\end{split}
\label{eq:sm_axial}
\end{equation}
and for each of the three basal bonds,
\begin{equation}
\begin{split}
s(In2\text{--}Se2) &= \exp\!\left[\frac{2.47 - 2.8508}{0.37}\right]\\
                   &= \exp(-1.0292) = 0.3573~\text{v.u.}
\end{split}
\label{eq:sm_basal}
\end{equation}
Summing the four contributions according to Eq.~(\ref{eq:sm_bvs}),
\begin{equation}
\sum s \,(Se2) = 0.8366 + 3 \times 0.3573 = 1.9085~\text{v.u.},
\label{eq:sm_sum}
\end{equation}
which is the value quoted in Table~II of the main text. The same procedure applied to the three equivalent $In1$--$Se1$ bonds at
$2.6259$~\AA\ gives $s=0.6562$~v.u. and $\sum s\,(Se1)=1.9685$~v.u., and to the
three equivalent $In2$--$Se3$ bonds at $2.6912$~\AA\ gives $s=0.5500$~v.u. and
$\sum s\,(Se3)=1.6500$~v.u. For Bi$_2$Se$_3$, with
$R_{0}(\text{Bi--Se})=2.72$~\AA, the three outer Se--Bi bonds at $2.8560$~\AA\
give $s=0.6924$~v.u. and $\sum s = 2.0771$~v.u., while the six central Se--Bi
bonds at $3.0791$~\AA\ give $s=0.3789$~v.u. and $\sum s = 2.2731$~v.u. All
values are collected in Table~\ref{tab:sm_bond_valence}.

The central result of this analysis is already visible in
Eqs.~(\ref{eq:sm_axial})--(\ref{eq:sm_sum}): although $Se2$ has one bond more
than $Se1$, three of its four bonds are the longest and therefore the weakest
In--Se bonds in the structure, so that the total bonding it carries,
$1.9085$~v.u., is in fact \emph{below} the $1.9685$~v.u.\ carried by the
three-coordinated $Se1$ site.

\subsection{Independence of the ratios from $R_{0}$}

Because $R_{0}$ enters Eq.~(\ref{eq:sm_bv}) only as an additive constant inside
the exponential, it cancels identically from any ratio of two bond valences of
the same element pair:
\begin{equation}
\frac{s(d_{1})}{s(d_{2})}
  = \frac{\exp[(R_{0}-d_{1})/b]}{\exp[(R_{0}-d_{2})/b]}
  = \exp\!\left[\frac{d_{2}-d_{1}}{b}\right],
\label{eq:sm_ratio}
\end{equation}
which depends only on the difference of the two bond lengths. The axial
$In1$--$Se2$ bond, for instance, is stronger than an $In1$--$Se1$ bond by
\begin{equation}
\frac{s(2.5360)}{s(2.6259)} = \exp\!\left[\frac{0.0899}{0.37}\right] = 1.275,
\label{eq:sm_ratio_num}
\end{equation}
independently of the value adopted for $R_{0}(\text{In--Se})$. The same holds
for the ratios $\sum s/\sum s_{\text{outer}}$ reported in the last column of
Table~II of the main text, since every bond entering those sums belongs to
the same element pair. We verified this numerically: varying
$R_{0}(\text{In--Se})$ between $2.40$ and $2.90$~\AA\ leaves
$\sum s\,(Se2)/\sum s\,(Se1) = 0.9695$ and
$\sum s\,(Se3)/\sum s\,(Se1) = 0.8382$ unchanged to six decimal places, while
the absolute sums are rescaled by a common factor of $3.86$. The ordering of the Se sites in
Table~II of the main text, and every comparison drawn from it in
Sec.~IV\,C, is therefore free of the one empirical parameter
of the model. Ratios formed \emph{across} the two compounds would not be, since
$R_{0}(\text{In--Se}) \ne R_{0}(\text{Bi--Se})$; for this reason
$\alpha$-In$_2$Se$_3$ and Bi$_2$Se$_3$ are each normalized to their own outer
Se site rather than compared directly.

\subsection{Accuracy of the absolute valence sums}

Applying Eq.~(\ref{eq:sm_bvs}) to the cations provides a consistency check on
the tabulated $R_{0}$. We obtain $V=2.805$~v.u.\ for the tetrahedral $In1$ and
$2.722$~v.u.\ for the octahedral $In2$, against a formal valence of $3$, and
$V=3.214$~v.u.\ for Bi, against the same formal value. The absolute sums are
thus accurate to roughly $7$--$10\%$, which is typical for selenides, where the
bond-valence parameters are determined from far fewer reference structures than
for oxides. The anion sums behave similarly: $1.968$, $1.909$ and $1.650$~v.u.\
for $Se1$, $Se2$ and $Se3$ against a formal $2$~v.u.
We therefore do not interpret the deviation of $Se3$ from $2$~v.u.\ as physical
underbonding, and use the bond valences only in the relative sense established
by Eq.~(\ref{eq:sm_ratio}).

\begin{table}[tb]
\caption{\label{tab:sm_bond_valence}%
Full-precision bond valences at every Se site of $\alpha$-In$_2$Se$_3$-3R and
Bi$_2$Se$_3$, from Eqs.~(\ref{eq:sm_bv}) and (\ref{eq:sm_bvs}) applied to the
relaxed r$^2$SCAN+rVV10 structures. Distances are means over
symmetry-equivalent bonds, which agree to within $2\times10^{-4}$~\AA.
$n$ is the number of bonds of that length at the site. Sums are evaluated at
full numerical precision, so the last digit of $\sum s$ may differ by one unit
from the product of the rounded entries shown.}
\begin{ruledtabular}
\begin{tabular}{lccccc}
Site & Neighbor & $n$ & $d$ (\AA) & $s$ (v.u.) & $\sum s$ (v.u.) \\
\colrule
\multicolumn{6}{l}{$\alpha$-In$_2$Se$_3$-3R, $R_{0}=2.47$~\AA}\\
$Se1$ & $In1$ & 3 & 2.6259 & 0.6562 & 1.9685 \\
$Se2$ & $In1$ & 1 & 2.5360 & 0.8366 & 1.9085 \\
      & $In2$ & 3 & 2.8508 & 0.3573 &        \\
$Se3$ & $In2$ & 3 & 2.6912 & 0.5500 & 1.6500 \\
\colrule
\multicolumn{6}{l}{Bi$_2$Se$_3$, $R_{0}=2.72$~\AA}\\
Se (outer)   & Bi & 3 & 2.8560 & 0.6924 & 2.0771 \\
Se (central) & Bi & 6 & 3.0791 & 0.3789 & 2.2731 \\
\end{tabular}
\end{ruledtabular}
\end{table}

\clearpage

\bibliography{bibliography}